\documentclass[a4paper,11pt]{article}
\pdfoutput=1
\usepackage{./auxi/jheppub}
\usepackage[utf8]{inputenc}
\usepackage{comment}

\usepackage{slashed} 
\usepackage{mathtools} 
\usepackage{tikz}
\usepackage{dsfont} 
\usetikzlibrary{positioning,arrows,calc}
\usetikzlibrary{arrows.meta}
\usetikzlibrary{decorations.pathmorphing}
\usetikzlibrary{decorations.markings}
\usetikzlibrary{shapes.geometric}
\usetikzlibrary{patterns}
\usepackage{xparse}
\usepackage{diagbox} 
\usepackage{tensor} 
\usepackage{subcaption} 

\newcommand{\Op}{\mathcal{O}}

\newcommand{\Wl}{\mathcal{W}_\ell}

\newcommand{\Wm}{\mathcal{W}}

\newcommand{\vev}[1]{\langle\, #1 \, \rangle}

\newif\ifstartcompletesineup
\newif\ifendcompletesineup
\pgfkeys{
    /pgf/decoration/.cd,
    start up/.is if=startcompletesineup,
    start up=true,
    start up/.default=true,
    start down/.style={/pgf/decoration/start up=false},
    end up/.is if=endcompletesineup,
    end up=true,
    end up/.default=true,
    end down/.style={/pgf/decoration/end up=false}
}
\pgfdeclaredecoration{complete sines}{initial}
{
    \state{initial}[
        width=+0pt,
        next state=upsine,
        persistent precomputation={
            \ifstartcompletesineup
                \pgfkeys{/pgf/decoration automaton/next state=upsine}
                \ifendcompletesineup
                    \pgfmathsetmacro\matchinglength{
                        0.5*\pgfdecoratedinputsegmentlength / (ceil(0.5* \pgfdecoratedinputsegmentlength / \pgfdecorationsegmentlength) )
                    }
                \else
                    \pgfmathsetmacro\matchinglength{
                        0.5 * \pgfdecoratedinputsegmentlength / (ceil(0.5 * \pgfdecoratedinputsegmentlength / \pgfdecorationsegmentlength ) - 0.499)
                    }
                \fi
            \else
                \pgfkeys{/pgf/decoration automaton/next state=downsine}
                \ifendcompletesineup
                    \pgfmathsetmacro\matchinglength{
                        0.5* \pgfdecoratedinputsegmentlength / (ceil(0.5 * \pgfdecoratedinputsegmentlength / \pgfdecorationsegmentlength ) - 0.4999)
                    }
                \else
                    \pgfmathsetmacro\matchinglength{
                        0.5 * \pgfdecoratedinputsegmentlength / (ceil(0.5 * \pgfdecoratedinputsegmentlength / \pgfdecorationsegmentlength ) )
                    }
                \fi
            \fi
            \setlength{\pgfdecorationsegmentlength}{\matchinglength pt}
        }] {}
    \state{downsine}[width=\pgfdecorationsegmentlength,next state=upsine]{
        \pgfpathsine{\pgfpoint{0.5\pgfdecorationsegmentlength}{0.5\pgfdecorationsegmentamplitude}}
        \pgfpathcosine{\pgfpoint{0.5\pgfdecorationsegmentlength}{-0.5\pgfdecorationsegmentamplitude}}
    }
    \state{upsine}[width=\pgfdecorationsegmentlength,next state=downsine]{
        \pgfpathsine{\pgfpoint{0.5\pgfdecorationsegmentlength}{-0.5\pgfdecorationsegmentamplitude}}
        \pgfpathcosine{\pgfpoint{0.5\pgfdecorationsegmentlength}{0.5\pgfdecorationsegmentamplitude}}
}
    \state{final}{}
}

\definecolor{bgbox}{RGB}{255,254,230}
\definecolor{setupplane}{RGB}{230,230,230}
\definecolor{gluoncolor}{RGB}{207,54,108}
\definecolor{vertexcolor}{RGB}{53,152,219}
\definecolor{SEcolor}{RGB}{176,156,255}
\definecolor{blobcolor}{RGB}{190,180,230}
\tikzset{
corner/.style={line width=1pt,dashed,draw=black,dash pattern=on 6pt off 4pt},
scalar/.style={line width=1pt,draw=black},
gluon/.style={line width=1pt,decorate, draw=gluoncolor,
    decoration={complete sines,aspect=0,amplitude=1.25mm,segment length=1.5mm,start up,end up}},
ghost/.style={line width=1pt,loosely dotted,draw=black},
wilson/.style={line width=2pt,draw=black},
 }
\NewDocumentCommand\semiloop{O{black}mmmO{}O{above}}
{%
\draw[#1] let \p1 = ($(#3)-(#2)$) in (#3) arc (#4:({#4+180}):({0.5*veclen(\x1,\y1)})node[midway, #6] {#5};)
}

\usepackage{bbm}

\newcommand\x{.6}
\newcommand\scale{.7}

\newcommand{\fourpttreeone}{\scalebox{\scale}{\begin{tikzpicture}[baseline={([yshift=-2ex]current bounding box.center)}]
\draw[wilson] (-2,0) -- (2,0);
\draw[scalar] (-1.6,0) arc (180:0:.7);
\draw[scalar] (.2,0) arc (180:0:.7);
\draw[fill=white] (-1.6,0) circle (.1);
\draw[fill=white] (-.2,0) circle (.1);
\draw[fill=white] (.2,0) circle (.1);
\draw[fill=white] (1.6,0) circle (.1);
\end{tikzpicture}}}
\newcommand{\fourpttreetwo}{\scalebox{\scale}{\begin{tikzpicture}[baseline={([yshift=-3.5ex]current bounding box.center)}]
\draw[wilson] (-2,0) -- (2,0);
\draw[scalar] (-1.6,0) arc (150:30:1.85);
\draw[scalar] (-.2,0) arc (210:-20:.23);
\draw[fill=white] (-1.6,0) circle (.1);
\draw[fill=white] (-.2,0) circle (.1);
\draw[fill=white] (.2,0) circle (.1);
\draw[fill=white] (1.6,0) circle (.1);
\end{tikzpicture}}}
\newcommand{\fourptSEone}{\scalebox{\scale}{\begin{tikzpicture}[baseline={([yshift=-2ex]current bounding box.center)}]
\draw[wilson] (-2,0) -- (2,0);
\draw[scalar] (-1.6,0) arc (180:0:.7);
\draw[scalar] (.2,0) arc (180:0:.7);
\draw[fill=white] (-1.6,0) circle (.1);
\draw[fill=white] (-.2,0) circle (.1);
\draw[fill=white] (.2,0) circle (.1);
\draw[fill=white] (1.6,0) circle (.1);
\draw[fill=SEcolor,SEcolor] (-.9,.7) circle (.15);
\end{tikzpicture}}}
\newcommand{\fourptSEtwo}{\scalebox{\scale}{\begin{tikzpicture}[baseline={([yshift=-2ex]current bounding box.center)}]
\draw[wilson] (-2,0) -- (2,0);
\draw[scalar] (-1.6,0) arc (180:0:.7);
\draw[scalar] (.2,0) arc (180:0:.7);
\draw[fill=white] (-1.6,0) circle (.1);
\draw[fill=white] (-.2,0) circle (.1);
\draw[fill=white] (.2,0) circle (.1);
\draw[fill=white] (1.6,0) circle (.1);
\draw[fill=SEcolor,SEcolor] (.9,.7) circle (.15);
\end{tikzpicture}}}
\newcommand{\fourptSEthree}{\scalebox{\scale}{\begin{tikzpicture}[baseline={([yshift=-3ex]current bounding box.center)}]
\draw[wilson] (-2,0) -- (2,0);
\draw[scalar] (-1.6,0) arc (150:30:1.85);
\draw[scalar] (-.2,0) arc (210:-20:.23);
\draw[fill=white] (-1.6,0) circle (.1);
\draw[fill=white] (-.2,0) circle (.1);
\draw[fill=white] (.2,0) circle (.1);
\draw[fill=white] (1.6,0) circle (.1);
\draw[fill=SEcolor,SEcolor] (0,.9) circle (.15);
\end{tikzpicture}}}
\newcommand{\fourptSEfour}{\scalebox{\scale}{\begin{tikzpicture}[baseline={([yshift=-2ex]current bounding box.center)}]
\draw[wilson] (-2,0) -- (2,0);
\draw[scalar] (-1.6,0) arc (150:30:1.85);
\draw[scalar] (-.2,0) arc (210:-20:.23);
\draw[fill=white] (-1.6,0) circle (.1);
\draw[fill=white] (-.2,0) circle (.1);
\draw[fill=white] (.2,0) circle (.1);
\draw[fill=white] (1.6,0) circle (.1);
\draw[fill=SEcolor,SEcolor] (0,.33) circle (.15);
\end{tikzpicture}}}
\newcommand{\fourptX}{\scalebox{\scale}{\begin{tikzpicture}[baseline={([yshift=-2ex]current bounding box.center)}]
\draw[wilson] (-2,0) -- (2,0);
\draw[scalar] (-1.6,0) arc (180:0:.9);
\draw[scalar] (-.2,0) arc (180:0:.9);
\draw[fill=white] (-1.6,0) circle (.1);
\draw[fill=white] (-.2,0) circle (.1);
\draw[fill=white] (.2,0) circle (.1);
\draw[fill=white] (1.6,0) circle (.1);
\draw[fill=vertexcolor,vertexcolor] (0,.55) circle (.075);
\end{tikzpicture}}}
\newcommand{\fourptHone}{\scalebox{\scale}{\begin{tikzpicture}[baseline={([yshift=-3ex]current bounding box.center)}]
\draw[wilson] (-2,0) -- (2,0);
\draw[scalar] (-1.6,0) arc (180:0:.7);
\draw[scalar] (.2,0) arc (180:0:.7);
\draw[gluon] (-.8,.68) arc (145:20:.95);
\draw[fill=white] (-1.6,0) circle (.1);
\draw[fill=white] (-.2,0) circle (.1);
\draw[fill=white] (.2,0) circle (.1);
\draw[fill=white] (1.6,0) circle (.1);
\draw[fill=vertexcolor,vertexcolor] (-.8,.68) circle (.075);
\draw[fill=vertexcolor,vertexcolor] (.8,.68) circle (.075);
\end{tikzpicture}}}
\newcommand{\fourptHtwo}{\scalebox{\scale}{\begin{tikzpicture}[baseline={([yshift=-3ex]current bounding box.center)}]
\draw[wilson] (-2,0) -- (2,0);
\draw[scalar] (-1.6,0) arc (150:30:1.85);
\draw[scalar] (-.2,0) arc (210:-20:.23);
\draw[gluon] (0,.3) -- (0,.9);
\draw[fill=white] (-1.6,0) circle (.1);
\draw[fill=white] (-.2,0) circle (.1);
\draw[fill=white] (.2,0) circle (.1);
\draw[fill=white] (1.6,0) circle (.1);
\draw[fill=vertexcolor,vertexcolor] (0,.91) circle (.075);
\draw[fill=vertexcolor,vertexcolor] (0,.33) circle (.075);
\end{tikzpicture}}}
\newcommand{\fourptYone}{\scalebox{\scale}{\begin{tikzpicture}[baseline={([yshift=-2.7ex]current bounding box.center)}]
\draw[wilson] (-2,0) -- (2,0);
\draw[scalar] (-1.6,0) arc (180:0:.7);
\draw[scalar] (.2,0) arc (180:0:.7);
\draw[gluon] (-.9,.7) -- (-.9,1.1);
\draw[fill=white] (-1.6,0) circle (.1);
\draw[fill=white] (-.2,0) circle (.1);
\draw[fill=white] (.2,0) circle (.1);
\draw[fill=white] (1.6,0) circle (.1);
\draw[fill=gluoncolor,gluoncolor] (-1.85,0) circle (.08);
\draw[fill=gluoncolor,gluoncolor] (-.9,0) circle (.08);
\draw[fill=gluoncolor,gluoncolor] (0,0) circle (.08);
\draw[fill=gluoncolor,gluoncolor] (1.85,0) circle (.08);
\draw[fill=vertexcolor,vertexcolor] (-.9,.7) circle (.075);
\end{tikzpicture}}}
\newcommand{\fourptYtwo}{\scalebox{\scale}{\begin{tikzpicture}[baseline={([yshift=-2.7ex]current bounding box.center)}]
\draw[wilson] (-2,0) -- (2,0);
\draw[scalar] (-1.6,0) arc (180:0:.7);
\draw[scalar] (.2,0) arc (180:0:.7);
\draw[gluon] (.9,.7) -- (.9,1.1);
\draw[fill=white] (-1.6,0) circle (.1);
\draw[fill=white] (-.2,0) circle (.1);
\draw[fill=white] (.2,0) circle (.1);
\draw[fill=white] (1.6,0) circle (.1);
\draw[fill=gluoncolor,gluoncolor] (-1.85,0) circle (.08);
\draw[fill=gluoncolor,gluoncolor] (0,0) circle (.08);
\draw[fill=gluoncolor,gluoncolor] (.9,0) circle (.08);
\draw[fill=gluoncolor,gluoncolor] (1.85,0) circle (.08);
\draw[fill=vertexcolor,vertexcolor] (.9,.7) circle (.075);
\end{tikzpicture}}}
\newcommand{\fourptYthree}{\scalebox{\scale}{\begin{tikzpicture}[baseline={([yshift=-3ex]current bounding box.center)}]
\draw[wilson] (-2,0) -- (2,0);
\draw[scalar] (-1.6,0) arc (150:30:1.85);
\draw[scalar] (-.2,0) arc (210:-20:.23);
\draw[gluon] (0,.9) -- (0,1.3);
\draw[fill=white] (-1.6,0) circle (.1);
\draw[fill=white] (-.2,0) circle (.1);
\draw[fill=white] (.2,0) circle (.1);
\draw[fill=white] (1.6,0) circle (.1);
\draw[fill=gluoncolor,gluoncolor] (-1.85,0) circle (.08);
\draw[fill=gluoncolor,gluoncolor] (-.9,0) circle (.08);
\draw[fill=gluoncolor,gluoncolor] (.9,0) circle (.08);
\draw[fill=gluoncolor,gluoncolor] (1.85,0) circle (.08);
\draw[fill=vertexcolor,vertexcolor] (0,.9) circle (.075);
\end{tikzpicture}}}
\newcommand{\fourptYfour}{\scalebox{\scale}{\begin{tikzpicture}[baseline={([yshift=-2ex]current bounding box.center)}]
\draw[wilson] (-2,0) -- (2,0);
\draw[scalar] (-1.6,0) arc (150:30:1.85);
\draw[scalar] (-.2,0) arc (210:-20:.23);
\draw[gluon] (0,.33) -- (0,.73);
\draw[fill=white] (-1.6,0) circle (.1);
\draw[fill=white] (-.2,0) circle (.1);
\draw[fill=white] (.2,0) circle (.1);
\draw[fill=white] (1.6,0) circle (.1);
\draw[fill=gluoncolor,gluoncolor] (-.9,0) circle (.08);
\draw[fill=gluoncolor,gluoncolor] (0,0) circle (.08);
\draw[fill=gluoncolor,gluoncolor] (.9,0) circle (.08);
\draw[fill=vertexcolor,vertexcolor] (0,.33) circle (.075);
\end{tikzpicture}}}

\newcommand{\recursiontree}{\scalebox{\scale}{\begin{tikzpicture}[baseline={([yshift=0ex]current bounding box.center)}]
\draw[wilson] (-2,0) -- (2,0);
\draw[scalar] (-1.5,0) arc (180:0:1);
\draw[scalar] (-.8,.3) arc (180:0:.3);
\draw[scalar] (-.8,0) -- (-.8,.3);
\draw[scalar] (-.2,0) -- (-.2,.3);
\draw[scalar] (.95,.3) arc (180:0:.3);
\draw[scalar] (.95,0) -- (.95,.3);
\draw[scalar] (1.55,0) -- (1.55,.3);
\draw[fill=white] (-1.5,0) circle (.1);
\draw[fill=white] (.5,0) circle (.1);
\node[] at (-1.5,-.4) {\small $1$};
\node[] at (.5,-.4) {\small $2j+2$};
\node[] at (-.5,.25) {\small $t$};
\node[] at (1.25,.25) {\small $t$};
\end{tikzpicture}}}
\newcommand{\treelevelblob}{\scalebox{\scale}{\begin{tikzpicture}[baseline={([yshift=-1.7ex]current bounding box.center)}]
\draw[wilson] (-1,0) -- (0,0);
\draw[scalar] (-.8,.3) arc (180:0:.3);
\draw[scalar] (-.8,0) -- (-.8,.3);
\draw[scalar] (-.2,0) -- (-.2,.3);
\node[] at (-.5,.25) {\small $t$};
\end{tikzpicture}}}
\newcommand{\recursionFour}{\scalebox{\scale}{\begin{tikzpicture}[baseline={([yshift=-1ex]current bounding box.center)}]
\draw[wilson] (-3,0) -- (3,0);
\draw[scalar] (-1.875,0) arc (180:0:1.25);
\draw[scalar] (-.625,0) arc (180:0:1.25);
\draw[scalar] (-2.8,.3) arc (180:0:.3);
\draw[scalar] (-2.8,0) -- (-2.8,.3);
\draw[scalar] (-2.2,0) -- (-2.2,.3);
\draw[scalar] (-1.55,.3) arc (180:0:.3);
\draw[scalar] (-1.55,0) -- (-1.55,.3);
\draw[scalar] (-.95,0) -- (-.95,.3);
\draw[scalar] (-.3,.3) arc (180:0:.3);
\draw[scalar] (-.3,0) -- (-.3,.3);
\draw[scalar] (.3,0) -- (.3,.3);
\draw[scalar] (.95,.3) arc (180:0:.3);
\draw[scalar] (.95,0) -- (.95,.3);
\draw[scalar] (1.55,0) -- (1.55,.3);
\draw[scalar] (2.2,.3) arc (180:0:.3);
\draw[scalar] (2.2,0) -- (2.2,.3);
\draw[scalar] (2.8,0) -- (2.8,.3);
\draw[fill=white] (-1.875,0) circle (.1);
\draw[fill=white] (-.625,0) circle (.1);
\draw[fill=white] (.625,0) circle (.1);
\draw[fill=white] (1.875,0) circle (.1);
\node[] at (-2.5,.25) {\small $t$};
\node[] at (-1.25,.25) {\small $t$};
\node[] at (0,.25) {\small $t$};
\node[] at (1.25,.25) {\small $t$};
\node[] at (2.5,.25) {\small $t$};
\node[] at (-1.875,-.4) {\small $i$};
\node[] at (-.625,-.4) {\small $j$};
\node[] at (.625,-.4) {\small $k$};
\node[] at (1.875,-.4) {\small $l$};
\draw[fill=white] (0,1.1) circle (.35);
\node[] at (0,1.1) {\small $4$};
\end{tikzpicture}}}

\newcommand{\recursionSE}{\scalebox{\scale}{\begin{tikzpicture}[baseline={([yshift=-1ex]current bounding box.center)}]
\draw[wilson] (-2.5,0) -- (2.5,0);
\draw[scalar] (-.75,.3) arc (180:0:.75);
\draw[scalar] (-.75,0) -- (-.75,.3);
\draw[scalar] (.75,0) -- (.75,.3);
\draw[scalar] (-1.8,.3) arc (180:0:.3);
\draw[scalar] (-1.8,0) -- (-1.8,.3);
\draw[scalar] (-1.2,0) -- (-1.2,.3);
\draw[scalar] (-.3,.3) arc (180:0:.3);
\draw[scalar] (-.3,0) -- (-.3,.3);
\draw[scalar] (.3,0) -- (.3,.3);
\draw[scalar] (1.2,.3) arc (180:0:.3);
\draw[scalar] (1.2,0) -- (1.2,.3);
\draw[scalar] (1.8,0) -- (1.8,.3);
\draw[fill=white] (-.75,0) circle (.1);
\draw[fill=white] (.75,0) circle (.1);
\draw[fill=SEcolor,SEcolor] (0,1.025) circle (.15);
\node[] at (-1.5,.25) {\small $t$};
\node[] at (0,.25) {\small $t$};
\node[] at (1.5,.25) {\small $t$};
\node[] at (-.75,-.4) {\small $i$};
\node[] at (.75,-.4) {\small $j$};
\end{tikzpicture}}}

\newcommand{\recursionYNone}{\scalebox{\scale}{\begin{tikzpicture}[baseline={([yshift=-2ex]current bounding box.center)}]
\draw[wilson] (-3.5,0) -- (2.5,0);
\draw[scalar] (-.75,.3) arc (180:0:.75);
\draw[scalar] (-.75,0) -- (-.75,.3);
\draw[scalar] (.75,0) -- (.75,.3);
\draw[gluon] (0,1.05) .. controls (-.5,1.6) and (-1.7,1.6)  .. (-2.1,-.1);
\draw[scalar] (-3,.3) arc (180:0:.3);
\draw[scalar] (-3,0) -- (-3,.3);
\draw[scalar] (-2.4,0) -- (-2.4,.3);
\draw[scalar] (-1.8,.3) arc (180:0:.3);
\draw[scalar] (-1.8,0) -- (-1.8,.3);
\draw[scalar] (-1.2,0) -- (-1.2,.3);
\draw[scalar] (-.3,.3) arc (180:0:.3);
\draw[scalar] (-.3,0) -- (-.3,.3);
\draw[scalar] (.3,0) -- (.3,.3);
\draw[scalar] (1.2,.3) arc (180:0:.3);
\draw[scalar] (1.2,0) -- (1.2,.3);
\draw[scalar] (1.8,0) -- (1.8,.3);
\draw[fill=white] (-2.4,0) circle (.1);
\draw[fill=white] (-.75,0) circle (.1);
\draw[fill=white] (.75,0) circle (.1);
\draw[fill=gluoncolor,gluoncolor] (-2.1,0) circle (.08);
\draw[fill=vertexcolor,vertexcolor] (0,1.05) circle (.075);
\node[] at (-2.7,.25) {\small $t$};
\node[] at (-1.5,.25) {\small $t$};
\node[] at (0,.25) {\small $t$};
\node[] at (1.5,.25) {\small $t$};
\node[] at (-.75,-.4) {\small $i$};
\node[] at (.75,-.4) {\small $j$};
\node[] at (-2.4,-.4) {\small $k$};
\end{tikzpicture}}}
\newcommand{\recursionYNtwo}{\scalebox{\scale}{\begin{tikzpicture}[baseline={([yshift=-2ex]current bounding box.center)}]
\draw[wilson] (-3,0) -- (3,0);
\draw[scalar] (-1.4,0) arc (165:16:1.45);
\draw[gluon] (0,1.05) -- (0,0);
\draw[scalar] (-2.5,.3) arc (180:0:.3);
\draw[scalar] (-2.5,0) -- (-2.5,.3);
\draw[scalar] (-1.9,0) -- (-1.9,.3);
\draw[scalar] (-.9,.3) arc (180:0:.3);
\draw[scalar] (-.9,0) -- (-.9,.3);
\draw[scalar] (-.3,0) -- (-.3,.3);
\draw[scalar] (.3,.3) arc (180:0:.3);
\draw[scalar] (.3,0) -- (.3,.3);
\draw[scalar] (.9,0) -- (.9,.3);
\draw[scalar] (1.9,.3) arc (180:0:.3);
\draw[scalar] (1.9,0) -- (1.9,.3);
\draw[scalar] (2.5,0) -- (2.5,.3);
\draw[fill=white] (-1.4,0) circle (.1);
\draw[fill=white] (-.3,0) circle (.1);
\draw[fill=white] (1.4,0) circle (.1);
\draw[fill=gluoncolor,gluoncolor] (0,0) circle (.08);
\draw[fill=vertexcolor,vertexcolor] (0,1.05) circle (.075);
\node[] at (-2.2,.25) {\small $t$};
\node[] at (-.6,.25) {\small $t$};
\node[] at (.6,.25) {\small $t$};
\node[] at (2.2,.25) {\small $t$};
\node[] at (-1.4,-.4) {\small $i$};
\node[] at (1.4,-.4) {\small $j$};
\node[] at (-.3,-.4) {\small $k$};
\end{tikzpicture}}}
\newcommand{\recursionYNthree}{\scalebox{\scale}{\begin{tikzpicture}[baseline={([yshift=-2ex]current bounding box.center)}]
\draw[wilson] (-2.5,0) -- (3.5,0);
\draw[scalar] (-.75,.3) arc (180:0:.75);
\draw[scalar] (-.75,0) -- (-.75,.3);
\draw[scalar] (.75,0) -- (.75,.3);
\draw[gluon] (0,1.05) .. controls (.5,1.6) and (1.7,1.6)  .. (2.1,-.1);
\draw[scalar] (-1.8,.3) arc (180:0:.3);
\draw[scalar] (-1.8,0) -- (-1.8,.3);
\draw[scalar] (-1.2,0) -- (-1.2,.3);
\draw[scalar] (-.3,.3) arc (180:0:.3);
\draw[scalar] (-.3,0) -- (-.3,.3);
\draw[scalar] (.3,0) -- (.3,.3);
\draw[scalar] (1.2,.3) arc (180:0:.3);
\draw[scalar] (1.2,0) -- (1.2,.3);
\draw[scalar] (1.8,0) -- (1.8,.3);
\draw[scalar] (2.4,.3) arc (180:0:.3);
\draw[scalar] (2.4,0) -- (2.4,.3);
\draw[scalar] (3,0) -- (3,.3);
\draw[fill=white] (1.8,0) circle (.1);
\draw[fill=white] (-.75,0) circle (.1);
\draw[fill=white] (.75,0) circle (.1);
\draw[fill=gluoncolor,gluoncolor] (2.1,0) circle (.08);
\draw[fill=vertexcolor,vertexcolor] (0,1.05) circle (.075);
\node[] at (2.7,.25) {\small $t$};
\node[] at (-1.5,.25) {\small $t$};
\node[] at (0,.25) {\small $t$};
\node[] at (1.5,.25) {\small $t$};
\node[] at (-.75,-.4) {\small $i$};
\node[] at (.75,-.4) {\small $j$};
\node[] at (1.8,-.4) {\small $k$};
\end{tikzpicture}}}
\newcommand{\recursionYexplainone}{\scalebox{\scale}{\begin{tikzpicture}[baseline={([yshift=-2ex]current bounding box.center)}]
\draw[wilson] (-3.5,0) -- (2.5,0);
\draw[scalar] (-.75,.3) arc (180:0:.75);
\draw[scalar] (-.75,0) -- (-.75,.3);
\draw[scalar] (.75,0) -- (.75,.3);
\draw[gluon] (0,1.05) .. controls (-.5,1.6) and (-1.7,1.6)  .. (-2.1,-.1);
\draw[scalar] (-3,.3) arc (180:0:.3);
\draw[scalar] (-3,0) -- (-3,.3);
\draw[scalar] (-2.4,0) -- (-2.4,.3);
\draw[scalar] (-1.8,.3) arc (180:0:.3);
\draw[scalar] (-1.8,0) -- (-1.8,.3);
\draw[scalar] (-1.2,0) -- (-1.2,.3);
\draw[scalar] (-.3,.3) arc (180:0:.3);
\draw[scalar] (-.3,0) -- (-.3,.3);
\draw[scalar] (.3,0) -- (.3,.3);
\draw[scalar] (1.2,.3) arc (180:0:.3);
\draw[scalar] (1.2,0) -- (1.2,.3);
\draw[scalar] (1.8,0) -- (1.8,.3);
\draw[fill=white] (-2.4,0) circle (.1);
\draw[fill=white] (-1.8,0) circle (.1);
\draw[fill=white] (-.75,0) circle (.1);
\draw[fill=white] (.75,0) circle (.1);
\draw[fill=gluoncolor,gluoncolor] (-2.1,0) circle (.08);
\draw[fill=vertexcolor,vertexcolor] (0,1.05) circle (.075);
\node[] at (-2.7,.25) {\small $t$};
\node[] at (-1.5,.25) {\small $t$};
\node[] at (0,.25) {\small $t$};
\node[] at (1.5,.25) {\small $t$};
\node[] at (-2.4,-.4) {\small $k$};
\node[] at (-1.6,-.4) {\small $k+1$};
\end{tikzpicture}}}
\newcommand{\recursionYexplaintwo}{\scalebox{\scale}{\begin{tikzpicture}[baseline={([yshift=-2ex]current bounding box.center)}]
\draw[wilson] (-3.5,0) -- (2.5,0);
\draw[scalar] (-.75,.3) arc (180:0:.75);
\draw[scalar] (-.75,0) -- (-.75,.3);
\draw[scalar] (.75,0) -- (.75,.3);
\draw[gluon] (0,1.05) .. controls (-.5,1.6) and (-1.7,1.6)  .. (-2.1,-.1);
\draw[scalar] (-3,.3) arc (180:0:.3);
\draw[scalar] (-3,0) -- (-3,.3);
\draw[scalar] (-2.4,0) -- (-2.4,.3);
\draw[scalar] (-1.8,.3) arc (180:0:.3);
\draw[scalar] (-1.8,0) -- (-1.8,.3);
\draw[scalar] (-1.2,0) -- (-1.2,.3);
\draw[scalar] (-.3,.3) arc (180:0:.3);
\draw[scalar] (-.3,0) -- (-.3,.3);
\draw[scalar] (.3,0) -- (.3,.3);
\draw[scalar] (1.2,.3) arc (180:0:.3);
\draw[scalar] (1.2,0) -- (1.2,.3);
\draw[scalar] (1.8,0) -- (1.8,.3);
\draw[fill=white] (-2.4,0) circle (.1);
\draw[fill=white] (-1.8,0) circle (.1);
\draw[fill=white] (-.75,0) circle (.1);
\draw[fill=white] (.75,0) circle (.1);
\draw[fill=gluoncolor,gluoncolor] (-2.1,0) circle (.08);
\draw[fill=vertexcolor,vertexcolor] (0,1.05) circle (.075);
\node[] at (-2.7,.25) {\small $t$};
\node[] at (-1.5,.25) {\small $t$};
\node[] at (0,.25) {\small $t$};
\node[] at (1.5,.25) {\small $t$};
\node[] at (-2.5,-.4) {\small $k$};
\node[gluoncolor] at (-2.1,-.4) {\small $\alpha$};
\node[] at (-1.4,-.4) {\small $k+1$};
\end{tikzpicture}}}
\newcommand{\recursionYLidone}{\scalebox{\scale}{\begin{tikzpicture}[baseline={([yshift=-2ex]current bounding box.center)}]
\draw[wilson] (-2.5,0) -- (2.5,0);
\draw[scalar] (-.75,.3) arc (180:0:.75);
\draw[scalar] (-.75,0) -- (-.75,.3);
\draw[scalar] (.75,0) -- (.75,.3);
\draw[gluon] (0,1.05) -- (0,1.5);
\draw[scalar] (-1.8,.3) arc (180:0:.3);
\draw[scalar] (-1.8,0) -- (-1.8,.3);
\draw[scalar] (-1.2,0) -- (-1.2,.3);
\draw[scalar] (-.3,.3) arc (180:0:.3);
\draw[scalar] (-.3,0) -- (-.3,.3);
\draw[scalar] (.3,0) -- (.3,.3);
\draw[scalar] (1.2,.3) arc (180:0:.3);
\draw[scalar] (1.2,0) -- (1.2,.3);
\draw[scalar] (1.8,0) -- (1.8,.3);
\draw[fill=white] (-.75,0) circle (.1);
\draw[fill=white] (.75,0) circle (.1);
\draw[fill=gluoncolor,gluoncolor] (-2.15,0) circle (.08);
\draw[fill=gluoncolor,gluoncolor] (-1.5,0) circle (.08);
\draw[fill=gluoncolor,gluoncolor] (-1,0) circle (.08);
\draw[fill=vertexcolor,vertexcolor] (0,1.05) circle (.075);
\node[] at (-1.5,.25) {\small $t$};
\node[] at (0,.25) {\small $t$};
\node[] at (1.5,.25) {\small $t$};
\node[] at (-.75,-.4) {\small $i$};
\node[] at (.75,-.4) {\small $j$};
\end{tikzpicture}}}

\newcommand{\recursionYone}{\scalebox{\scale}{\begin{tikzpicture}[baseline={([yshift=-2ex]current bounding box.center)}]
\draw[wilson] (-2.5,0) -- (2.5,0);
\draw[scalar] (-.75,.3) arc (180:0:.75);
\draw[scalar] (-.75,0) -- (-.75,.3);
\draw[scalar] (.75,0) -- (.75,.3);
\draw[gluon] (0,1.05) -- (0,1.5);
\draw[scalar] (-1.8,.3) arc (180:0:.3);
\draw[scalar] (-1.8,0) -- (-1.8,.3);
\draw[scalar] (-1.2,0) -- (-1.2,.3);
\draw[scalar] (-.3,.3) arc (180:0:.3);
\draw[scalar] (-.3,0) -- (-.3,.3);
\draw[scalar] (.3,0) -- (.3,.3);
\draw[scalar] (1.2,.3) arc (180:0:.3);
\draw[scalar] (1.2,0) -- (1.2,.3);
\draw[scalar] (1.8,0) -- (1.8,.3);
\draw[fill=white] (-.75,0) circle (.1);
\draw[fill=white] (.75,0) circle (.1);
\draw[fill=gluoncolor,gluoncolor] (-2.15,0) circle (.08);
\draw[fill=gluoncolor,gluoncolor] (-1.5,0) circle (.08);
\draw[fill=gluoncolor,gluoncolor] (-1,0) circle (.08);
\draw[fill=gluoncolor,gluoncolor] (-.5,0) circle (.08);
\draw[fill=gluoncolor,gluoncolor] (0,0) circle (.08);
\draw[fill=gluoncolor,gluoncolor] (.5,0) circle (.08);
\draw[fill=gluoncolor,gluoncolor] (1,0) circle (.08);
\draw[fill=gluoncolor,gluoncolor] (1.5,0) circle (.08);
\draw[fill=gluoncolor,gluoncolor] (2.15,0) circle (.08);
\draw[fill=vertexcolor,vertexcolor] (0,1.05) circle (.075);
\node[] at (-1.5,.25) {\small $t$};
\node[] at (0,.25) {\small $t$};
\node[] at (1.5,.25) {\small $t$};
\node[] at (-.75,-.4) {\small $i$};
\node[] at (.75,-.4) {\small $j$};
\end{tikzpicture}}}

\newcommand{\recursionYLtwo}{\scalebox{\scale}{\begin{tikzpicture}[baseline={([yshift=-2ex]current bounding box.center)}]
\draw[wilson] (-5,0) -- (2.5,0);
\draw[scalar] (-.75,.3) arc (180:0:.75);
\draw[scalar] (-.75,0) -- (-.75,.3);
\draw[scalar] (.75,0) -- (.75,.3);
\draw[gluon] (0,1.05) -- (0,1.5);
\draw[scalar] (-3.75,.3) arc (180:0:.75);
\draw[scalar] (-3.75,0) -- (-3.75,.3);
\draw[scalar] (-2.25,0) -- (-2.25,.3);
\draw[scalar] (-4.8,.3) arc (180:0:.3);
\draw[scalar] (-4.8,0) -- (-4.8,.3);
\draw[scalar] (-4.2,0) -- (-4.2,.3);
\draw[scalar] (-3.3,.3) arc (180:0:.3);
\draw[scalar] (-3.3,0) -- (-3.3,.3);
\draw[scalar] (-2.7,0) -- (-2.7,.3);
\draw[scalar] (-1.8,.3) arc (180:0:.3);
\draw[scalar] (-1.8,0) -- (-1.8,.3);
\draw[scalar] (-1.2,0) -- (-1.2,.3);
\draw[scalar] (-.3,.3) arc (180:0:.3);
\draw[scalar] (-.3,0) -- (-.3,.3);
\draw[scalar] (.3,0) -- (.3,.3);
\draw[scalar] (1.2,.3) arc (180:0:.3);
\draw[scalar] (1.2,0) -- (1.2,.3);
\draw[scalar] (1.8,0) -- (1.8,.3);
\draw[fill=white] (-.75,0) circle (.1);
\draw[fill=white] (.75,0) circle (.1);
\draw[fill=white] (-3.75,0) circle (.1);
\draw[fill=white] (-2.25,0) circle (.1);
\draw[fill=gluoncolor,gluoncolor] (-3.5,0) circle (.08);
\draw[fill=gluoncolor,gluoncolor] (-2.5,0) circle (.08);
\draw[fill=gluoncolor,gluoncolor] (-3,0) circle (.08);
\draw[fill=vertexcolor,vertexcolor] (0,1.05) circle (.075);
\node[] at (-4.5,.25) {\small $t$};
\node[] at (-3,.25) {\small $t$};
\node[] at (-1.5,.25) {\small $t$};
\node[] at (0,.25) {\small $t$};
\node[] at (1.5,.25) {\small $t$};
\node[] at (-3.75,-.4) {\small $\alpha$};
\node[] at (-2.25,-.4) {\small $\beta$};
\node[] at (-.75,-.4) {\small $i$};
\node[] at (.75,-.4) {\small $j$};
\end{tikzpicture}}}

\newcommand{\recursionYRtwo}{\scalebox{\scale}{\begin{tikzpicture}[baseline={([yshift=-2ex]current bounding box.center)}]
\draw[wilson] (-2.5,0) -- (5,0);
\draw[scalar] (-.75,.3) arc (180:0:.75);
\draw[scalar] (-.75,0) -- (-.75,.3);
\draw[scalar] (.75,0) -- (.75,.3);
\draw[gluon] (0,1.05) -- (0,1.5);
\draw[scalar] (-1.8,.3) arc (180:0:.3);
\draw[scalar] (-1.8,0) -- (-1.8,.3);
\draw[scalar] (-1.2,0) -- (-1.2,.3);
\draw[scalar] (-.3,.3) arc (180:0:.3);
\draw[scalar] (-.3,0) -- (-.3,.3);
\draw[scalar] (.3,0) -- (.3,.3);
\draw[scalar] (1.2,.3) arc (180:0:.3);
\draw[scalar] (1.2,0) -- (1.2,.3);
\draw[scalar] (1.8,0) -- (1.8,.3);
\draw[scalar] (2.7,.3) arc (180:0:.3);
\draw[scalar] (2.7,0) -- (2.7,.3);
\draw[scalar] (3.3,0) -- (3.3,.3);
\draw[scalar] (4.2,.3) arc (180:0:.3);
\draw[scalar] (4.2,0) -- (4.2,.3);
\draw[scalar] (4.8,0) -- (4.8,.3);
\draw[scalar] (2.25,.3) arc (180:0:.75);
\draw[scalar] (2.25,0) -- (2.25,.3);
\draw[scalar] (3.75,0) -- (3.75,.3);
\draw[fill=white] (-.75,0) circle (.1);
\draw[fill=white] (.75,0) circle (.1);
\draw[fill=white] (2.25,0) circle (.1);
\draw[fill=white] (3.75,0) circle (.1);
\draw[fill=gluoncolor,gluoncolor] (2.5,0) circle (.08);
\draw[fill=gluoncolor,gluoncolor] (3,0) circle (.08);
\draw[fill=gluoncolor,gluoncolor] (3.5,0) circle (.08);
\draw[fill=vertexcolor,vertexcolor] (0,1.05) circle (.075);
\node[] at (-1.5,.25) {\small $t$};
\node[] at (0,.25) {\small $t$};
\node[] at (1.5,.25) {\small $t$};
\node[] at (3,.25) {\small $t$};
\node[] at (4.5,.25) {\small $t$};
\node[] at (-.75,-.4) {\small $i$};
\node[] at (.75,-.4) {\small $j$};
\node[] at (2.25,-.4) {\small $\alpha$};
\node[] at (3.75,-.4) {\small $\beta$};
\end{tikzpicture}}}
\newcommand{\recursionbridge}{\scalebox{\scale}{\begin{tikzpicture}[baseline={([yshift=-3ex]current bounding box.center)}]
\draw[wilson] (-3,0) -- (3,0);
\draw[scalar] (-1.875,0) arc (180:0:1.875);
\draw[scalar] (-1.25,0) arc (180:0:1.25);
\draw[scalar] (-2.8,.3) arc (180:0:.3);
\draw[scalar] (-2.8,0) -- (-2.8,.3);
\draw[scalar] (-2.2,0) -- (-2.2,.3);
\draw[scalar] (2.2,.3) arc (180:0:.3);
\draw[scalar] (2.2,0) -- (2.2,.3);
\draw[scalar] (2.8,0) -- (2.8,.3);
\draw[fill=white] (-1.875,0) circle (.1);
\draw[fill=white] (1.875,0) circle (.1);
\node[] at (-2.5,.25) {\small $t$};
\node[] at (2.5,.25) {\small $t$};
\node[] at (-1.875,-.4) {\small $i$};
\node[] at (1.875,-.4) {\small $j$};
\node[] at (0,.5) {$A_{j-i-1}^{\text{\tiny{NLO}}}$};
\end{tikzpicture}}}

\newcommand{\recursionYCtwodue}{\scalebox{\scale}{\begin{tikzpicture}[baseline={([yshift=-3.9ex]current bounding box.center)}]
\draw[wilson] (-3.75,0) -- (3.75,0);
\draw[scalar] (2.25,0) arc(0:180:2.25cm and 1.7cm);
\draw[scalar] (-.75,.3) arc (180:0:.75);
\draw[scalar] (-.75,0) -- (-.75,.3);
\draw[scalar] (.75,0) -- (.75,.3);
\draw[gluon] (0,1.7) -- (0,2.15);
\draw[scalar] (-3.3,.3) arc (180:0:.3);
\draw[scalar] (-3.3,0) -- (-3.3,.3);
\draw[scalar] (-2.7,0) -- (-2.7,.3);
\draw[scalar] (-1.8,.3) arc (180:0:.3);
\draw[scalar] (-1.8,0) -- (-1.8,.3);
\draw[scalar] (-1.2,0) -- (-1.2,.3);
\draw[scalar] (-.3,.3) arc (180:0:.3);
\draw[scalar] (-.3,0) -- (-.3,.3);
\draw[scalar] (.3,0) -- (.3,.3);
\draw[scalar] (1.2,.3) arc (180:0:.3);
\draw[scalar] (1.2,0) -- (1.2,.3);
\draw[scalar] (1.8,0) -- (1.8,.3);
\draw[scalar] (2.7,.3) arc (180:0:.3);
\draw[scalar] (2.7,0) -- (2.7,.3);
\draw[scalar] (3.3,0) -- (3.3,.3);
\draw[fill=white] (-.75,0) circle (.1);
\draw[fill=white] (.75,0) circle (.1);
\draw[fill=white] (-2.25,0) circle (.1);
\draw[fill=white] (2.25,0) circle (.1);
\draw[fill=gluoncolor,gluoncolor] (-.5,0) circle (.08);
\draw[fill=gluoncolor,gluoncolor] (0,0) circle (.08);
\draw[fill=gluoncolor,gluoncolor] (.5,0) circle (.08);
\draw[fill=vertexcolor,vertexcolor] (0,1.7) circle (.075);
\node[] at (-3,.25) {\small $t$};
\node[] at (-1.5,.25) {\small $t$};
\node[] at (0,.25) {\small $t$};
\node[] at (1.5,.25) {\small $t$};
\node[] at (3,.25) {\small $t$};
\node[] at (-.75,-.4) {\small $i$};
\node[] at (.75,-.4) {\small $j$};
\node[] at (-2.25,-.4) {\small $\alpha$};
\node[] at (2.25,-.4) {\small $\beta$};
\end{tikzpicture}}}

\newcommand{\propagatorS}{\begin{tikzpicture}[baseline={([yshift=-.35ex]current bounding box.center)}]
\draw[scalar] (0,.5) -- (1,.5);
\draw[fill=white] (0,.5) circle (.075);
\draw[fill=white] (1,.5) circle (.075);
\node[] at (0,.1) {\footnotesize \makebox[\widthof{$\smash{\mu,a}$}][c]{$\smash{i,a}$}};
\node[] at (0,.9) {\footnotesize $1$};
\node[] at (1,.1) {\footnotesize \makebox[\widthof{$\smash{\mu,a}$}][c]{$\smash{j,b}$}};
\node[] at (1,.9) {\footnotesize $2$};
\end{tikzpicture}}
\newcommand{\propagatorG}{\begin{tikzpicture}[baseline={([yshift=-.35ex]current bounding box.center)}]
\draw[gluon] (0,.5) -- (1,.5);
\draw[fill=white] (0,.5) circle (.075);
\draw[fill=white] (1,.5) circle (.075);
\node[] at (0,.1) {\footnotesize \makebox[\widthof{$\mu,a$}][c]{$\smash{\mu,a}$}};
\node[] at (0,.9) {\footnotesize $1$};
\node[] at (1,.1) {\footnotesize \makebox[\widthof{$\nu,b$}][c]{$\smash{\nu,b}$}};
\node[] at (1,.9) {\footnotesize $2$};
\end{tikzpicture}}
\newcommand{\propagatorF}{\begin{tikzpicture}[baseline={([yshift=-.35ex]current bounding box.center)}]
\draw[scalar] (0,.5) -- (1,.5);
\draw[fill=white] (0,.5) circle (.075);
\draw[fill=white] (1,.5) circle (.075);
\draw[-Latex] (.62,.5)--(.63,.5);
\node[] at (0,.1) {\footnotesize \makebox[\widthof{$\smash{\mu,a}$}][c]{$\smash{a}$}};
\node[] at (0,.9) {\footnotesize $1$};
\node[] at (1,.1) {\footnotesize \makebox[\widthof{$\smash{\nu,b}$}][c]{$\smash{b}$}};
\node[] at (1,.9) {\footnotesize $2$};
\end{tikzpicture}}
\newcommand{\propagatorGh}{\begin{tikzpicture}[baseline={([yshift=-.35ex]current bounding box.center)}]
\draw[ghost] (0,.5) -- (1,.5);
\draw[fill=white] (0,.5) circle (.075);
\draw[fill=white] (1,.5) circle (.075);
\node[] at (0,.1) {\footnotesize \makebox[\widthof{$\smash{\mu,a}$}][c]{$\smash{a}$}};
\node[] at (0,.9) {\footnotesize $1$};
\node[] at (1,.1) {\footnotesize \makebox[\widthof{$\smash{\nu,b}$}][c]{$\smash{b}$}};
\node[] at (1,.9) {\footnotesize $2$};
\end{tikzpicture}}

\newcommand{\propagatorSSEnotext}{\begin{tikzpicture}[baseline={([yshift=-.35ex]current bounding box.center)}]
\draw[scalar] (0,.5) -- (1,.5);
\draw[fill=white] (0,.5) circle (.075);
\draw[fill=white] (1,.5) circle (.075);
\draw[fill=SEcolor,SEcolor] (.5,.5) circle (.15);
\end{tikzpicture}}
\newcommand{\SSEone}{\begin{tikzpicture}[baseline={([yshift=-.4ex]current bounding box.center)}]
\draw[scalar] (0,.5) -- (2,.5);
\draw[gluon] (.5,.5) arc (180:-20:.5);
\draw[fill=white] (0,.5) circle (.075);
\draw[fill=white] (2,.5) circle (.075);
\draw[fill=vertexcolor,vertexcolor] (.5,.5) circle (.075);
\draw[fill=vertexcolor,vertexcolor] (1.5,.5) circle (.075);
\end{tikzpicture}}
\newcommand{\SSEtwo}{\begin{tikzpicture}[baseline={([yshift=-.5ex]current bounding box.center)}]
\draw[scalar] (0,.5) -- (.5,.5);
\draw[scalar] (1.5,.5) -- (2,.5);
\draw[scalar,postaction={decorate}] (.5,.5) arc (180:0:.5);
\draw[-Latex] (1.1,1)--(1.11,1);
\draw[scalar,postaction={decorate}] (.5,.5) arc (-180:0:.5);
\draw[-Latex] (.9,0)--(.89,0);
\draw[fill=white] (0,.5) circle (.075);
\draw[fill=white] (2,.5) circle (.075);
\draw[fill=vertexcolor,vertexcolor] (.5,.5) circle (.075);
\draw[fill=vertexcolor,vertexcolor] (1.5,.5) circle (.075);
\end{tikzpicture}}
\newcommand{\SSEthree}{\begin{tikzpicture}[baseline={([yshift=-.75ex]current bounding box.center)}]
\draw[scalar] (0,.5) -- (2,.5);
\draw[scalar] (1,1) circle (.5);
\draw[fill=white] (0,.5) circle (.075);
\draw[fill=white] (2,.5) circle (.075);
\draw[fill=vertexcolor,vertexcolor] (1,.5) circle (.075);
\end{tikzpicture}}
\newcommand{\SSEfour}{\begin{tikzpicture}[baseline={([yshift=-1.05ex]current bounding box.center)}]
\draw[scalar] (0,.5) -- (2,.5);
\draw[gluon] (1,1) circle (.5);
\draw[fill=white] (0,.5) circle (.075);
\draw[fill=white] (2,.5) circle (.075);
\draw[fill=vertexcolor,vertexcolor] (1,.5) circle (.075);
\end{tikzpicture}}
\newcommand{\vertexSSG}{\begin{tikzpicture}[baseline={([yshift=-.35ex]current bounding box.center)}]
\draw[gluon] (0,.5) -- (.75,.5);
\draw[scalar] (.75,.5) -- (1.25,1);
\draw[scalar] (.75,.5) -- (1.25,0);
\draw[fill=vertexcolor,vertexcolor] (.75,.5) circle (.075);
\draw[fill=white] (0,.5) circle (.075);
\draw[fill=white] (1.25,1) circle (.075);
\draw[fill=white] (1.25,0) circle (.075);
\node[anchor=west] at (1.3,1) {\footnotesize $i,a$};
\node[] at (1.25,1.3) {\footnotesize $1$};
\node[anchor=west] at (1.3,0) {\footnotesize $j,b$};
\node[] at (1.25,-.3) {\footnotesize $2$};
\node[] at (0,.1) {\footnotesize $\mu,c$};
\node[] at (0,.9) {\footnotesize $3$};
\end{tikzpicture}}
\newcommand{\vertexSSSS}{\begin{tikzpicture}[baseline={([yshift=-.35ex]current bounding box.center)}]
\draw[scalar] (0,0) -- (1.25,1);
\draw[scalar] (0,1) -- (1.25,0);
\draw[fill=vertexcolor,vertexcolor] (.625,.5) circle (.075);
\draw[fill=white] (0,0) circle (.075);
\draw[fill=white] (0,1) circle (.075);
\draw[fill=white] (1.25,1) circle (.075);
\draw[fill=white] (1.25,0) circle (.075);
\node[anchor=west] at (1.3,1) {\footnotesize $i,a$};
\node[] at (1.25,1.3) {\footnotesize $1$};
\node[anchor=west] at (1.3,0) {\footnotesize $j,b$};
\node[] at (1.25,-.3) {\footnotesize $2$};
\node[anchor=east] at (-0.05,1) {\footnotesize $k,c$};
\node[] at (0,1.3) {\footnotesize $3$};
\node[anchor=east] at (-0.05,0) {\footnotesize $l,d$};
\node[] at (0,-.3) {\footnotesize $4$};
\end{tikzpicture}}
\newcommand{\vertexSSSSno}{\begin{tikzpicture}[baseline={([yshift=-.35ex]current bounding box.center)}]
\draw[scalar] (0,0) -- (1.25,1);
\draw[scalar] (0,1) -- (1.25,0);
\draw[fill=vertexcolor,vertexcolor] (.625,.5) circle (.075);
\draw[fill=white] (0,0) circle (.075);
\draw[fill=white] (0,1) circle (.075);
\draw[fill=white] (1.25,1) circle (.075);
\draw[fill=white] (1.25,0) circle (.075);
\end{tikzpicture}}
\newcommand{\vertexFourno}{\begin{tikzpicture}[baseline={([yshift=-.35ex]current bounding box.center)}]
\draw[scalar] (0,0) -- (1.25,1);
\draw[scalar] (0,1) -- (1.25,0);
\draw[fill=white] (0,0) circle (.075);
\draw[fill=white] (0,1) circle (.075);
\draw[fill=white] (1.25,1) circle (.075);
\draw[fill=white] (1.25,0) circle (.075);
\draw[fill=white] (.625,.5) circle (.35);
\node[] at (.625,.5) {\small $4$};
\end{tikzpicture}}

\newcommand{\Hinsertone}{\begin{tikzpicture}[baseline={([yshift=-.55ex]current bounding box.center)}]
\draw[scalar] (0,0) -- (1.25,0);
\draw[scalar] (0,1) -- (1.25,1);
\draw[gluon] (.625,0) -- (.625,1);
\draw[fill=vertexcolor,vertexcolor] (.625,0) circle (.075);
\draw[fill=vertexcolor,vertexcolor] (.625,1) circle (.075);
\draw[fill=white] (0,0) circle (.075);
\draw[fill=white] (0,1) circle (.075);
\draw[fill=white] (1.25,1) circle (.075);
\draw[fill=white] (1.25,0) circle (.075);
\node[anchor=west] at (1.3,1) {\footnotesize $i,a$};
\node[] at (1.25,1.3) {\footnotesize $1$};
\node[anchor=west] at (1.3,0) {\footnotesize $j,b$};
\node[] at (1.25,-.3) {\footnotesize $2$};
\node[anchor=east] at (-0.05,1) {\footnotesize $k,c$};
\node[] at (0,1.3) {\footnotesize $3$};
\node[anchor=east] at (-0.05,0) {\footnotesize $l,d$};
\node[] at (0,-.3) {\footnotesize $4$};
\end{tikzpicture}}
\newcommand{\Hinsertoneno}{\begin{tikzpicture}[baseline={([yshift=-.55ex]current bounding box.center)}]
\draw[scalar] (0,0) -- (1.25,0);
\draw[scalar] (0,1) -- (1.25,1);
\draw[gluon] (.625,0) -- (.625,1);
\draw[fill=vertexcolor,vertexcolor] (.625,0) circle (.075);
\draw[fill=vertexcolor,vertexcolor] (.625,1) circle (.075);
\draw[fill=white] (0,0) circle (.075);
\draw[fill=white] (0,1) circle (.075);
\draw[fill=white] (1.25,1) circle (.075);
\draw[fill=white] (1.25,0) circle (.075);
\end{tikzpicture}}

\newcommand{\Hinserttwono}{\begin{tikzpicture}[baseline={([yshift=-.55ex]current bounding box.center)}]
\draw[gluon] (0,.5) -- (1.25,.5);
\draw[scalar] (1.25,0) -- (1.25,1);
\draw[scalar] (0,0) -- (0,1);
\draw[fill=vertexcolor,vertexcolor] (0,.5) circle (.075);
\draw[fill=vertexcolor,vertexcolor] (1.25,.5) circle (.075);
\draw[fill=white] (0,0) circle (.075);
\draw[fill=white] (0,1) circle (.075);
\draw[fill=white] (1.25,1) circle (.075);
\draw[fill=white] (1.25,0) circle (.075);
\end{tikzpicture}}

\newcommand{\diagYonetwotwo}{\scalebox{\x}{\begin{tikzpicture}[baseline={([yshift=-.55ex]current bounding box.center)}]
\draw[scalar] (0,.5) -- (1,.5);
\draw[scalar] (1.5,.5) circle (.5);
\draw[fill=white] (0,.5) circle (.075);
\draw[fill=white] (2,.5) circle (.075);
\draw[fill=vertexcolor,vertexcolor] (1,.5) circle (.075);
\end{tikzpicture}}}
\newcommand{\diagXoneonetwothree}{\scalebox{\x}{\begin{tikzpicture}[baseline={([yshift=-.55ex]current bounding box.center)}]
\draw[scalar] (.5,1) -- (1,.5);
\draw[scalar] (.5,0) -- (1,.5);
\draw[scalar] (1.5,.5) circle (.5);
\draw[fill=white] (.5,1) circle (.075);
\draw[fill=white] (.5,0) circle (.075);
\draw[fill=white] (2,.5) circle (.075);
\draw[fill=vertexcolor,vertexcolor] (1,.5) circle (.075);
\end{tikzpicture}}}

\usepackage{simplewick}
\usepackage{dsfont}
\usepackage{float}
\usepackage{pgfplots}
\pgfplotsset{compat=1.14}

\let\oldbfseries=\bfseries
\let\oldmdseries=\mdseries
\let\oldnormalfont=\normalfont
\renewcommand{\bfseries}{\oldbfseries\boldmath}
\renewcommand{\mdseries}{\oldmdseries\unboldmath}
\renewcommand{\normalfont}{\oldnormalfont\unboldmath}

\makeatletter
\newlength{\apb@width}
\newcommand{\autoparbox}[2][c]{\settowidth{\apb@width}{#2}\parbox[#1]{\apb@width}{#2}}

\makeatother

\DeclareMathOperator{\tr}{tr}

\def\Am{{\mathcal{A}}}

\def\Km{{\mathcal{K}}}

\def\Nm{{\mathcal{N}}}

\def\Pm{{\mathcal{P}}}

\newcommand{\beq}{\begin{equation}}
\newcommand{\eeq}{\end{equation}}

\definecolor{nicegreen}{rgb}{0.1,0.6,0.1}

\newmuskip\pFqskip
\mathchardef\pFcomma=\mathcode`,

\makeatletter
\renewcommand*\env@matrix[1][\arraystretch]{%
  \edef\arraystretch{#1}%
  \hskip -\arraycolsep
  \let\@ifnextchar\new@ifnextchar
  \array{*\c@MaxMatrixCols c}}
\makeatother

\graphicspath{{./figures/}}

\title{\center{Multipoint correlators on the supersymmetric \\ Wilson line defect CFT}}

\author[1]{Julien Barrat,}
\author[2]{Pedro Liendo,}
\author[1]{Giulia Peveri,}
\author[1]{Jan Plefka.}

\affiliation[1]{Institut f\"ur Physik und IRIS Adlershof, Humboldt-Universit{\"a}t zu Berlin, Zum Gro{\ss}en Windkanal 2, 12489 Berlin, Germany}
\affiliation[2]{DESY Hamburg, Theory Group, Notkestra{\ss}e 85, D-22607 Hamburg, Germany}

\emailAdd{julien.barrat@hu-berlin.de, pedro.liendo@desy.de, giulia.peveri@physik.hu-berlin.de,jan.plefka@hu-berlin.de}

\preprint{HU-EP-21/53-RTG}

\abstract{We study multipoint correlators of protected scalars on the 
Maldacena-Wilson line in $\Nm=4$ SYM. Working at weak coupling in the planar limit, we derive an explicit recursion relation that captures next-to-leading order correlators with an arbitrary number of insertions of the fundamental scalar field. By pinching fundamental scalars together, we can build composite protected operators with higher values of the $R$-charge. Our result then encompasses arbitrary $n$-point correlators of protected operators with arbitrary weight. As a demonstration of our method, we give explicit formulae for correlators with up to six points. Using these results we observe that all our correlators are annihilated by a special class of differential operators. We conjecture that these differential operators are non-perturbative constraints and can be considered a multipoint extension of the superconformal Ward identities satisfied by four-point functions.}

\begin{document} 

\setcounter{tocdepth}{2}
\maketitle
\setcounter{page}{1}

\section{Introduction}
\label{sec:intro}

There exist several modern approaches for the analytical study of quantum field theories. In addition to generalized unitarity techniques in perturbation theory, the past decades have seen developments that include holography, integrability, localization and the conformal bootstrap.  At the crossroads of these techniques is the maximally supersymmetric $\Nm=4$ super Yang-Mills (SYM) theory, which is considered to be the simplest interacting quantum field theory in four dimensions.  In particular, the 't Hooft large $N$ limit of the color gauge group $SU(N)$ is conjectured to be integrable, and an enormous amount of results has been obtained, including non-perturbative ones, which could be checked against the dual type IIB superstring in an AdS$_5 \times$ S$^5$ spacetime background. 

Apart from the spectrum of local operators and their correlation functions, an important 
non-local observable in gauge theories is the Wilson loop. This operator describes the coupling between a heavy (probe) particle and the gauge fields of the theory. In $\Nm = 4$ SYM one may consider supersymmetric extensions of this operator, known as Maldacena-Wilson loop operators \cite{Maldacena:1998im,sjrey}. These extend the usual Wilson loop operator coupling to the gauge fields $A_{\mu}(x)$ by a coupling of the adjoint scalar fields $\phi_{i}(x)$ to a path $C$ on $\mathbb{R}^{1,3}\times$ S$^{5}$:
\begin{equation}
\Wm_C := \frac{1}{N} \text{tr}\, \Pm \exp \int_C d\tau\, (\dot{x}_\mu A^\mu (x) + 
\sqrt{\dot x^{\mu}\, \dot x_{\mu}}\, \theta \cdot \phi (x))\,,
\label{eq:WC}
\end{equation}
where $\theta^{i}$ is an $SO(6)$ vector parametrizing a path on the $S^{5}$.  
The coupling is such that from a $10d$ perspective the path is light-like:  $\dot x^{M}=\{
\dot x^{\mu}, \theta^{i} \sqrt{\dot x^{\mu}\, \dot x_{\mu}}\}$ with $\dot x^{M} \dot x_{M}=0$
in mostly minus signature. This operator is locally half-BPS and conformally invariant.
Remarkably, for special geometries of the path $C$, such as the circle and the infinite straight line, the expectation value of this operator was obtained at all orders by summing up Feynman diagrams \cite{Erickson:2000af,Drukker:2000rr}; a result which was later confirmed rigorously using supersymmetric localization \cite{Pestun:2007rz,Pestun:2009nn}.\footnote{The complete supersymmetric extension of the Maldacena-Wilson loop was presented in
\cite{Muller:2013rta,Beisert:2015jxa,Beisert:2015uda} coupling to a path in non-chiral $\mathcal{N}=4$ superspace.} This is the natural dual object to the minimal type IIB superstring surface in the AdS/CFT correspondence and it enjoys a Yangian invariance. 

In recent years, there has been a renewal of interest in the Maldacena-Wilson loop from the point of view of conformal defects, i.e. extended operators which preserve some of the original conformal symmetry. In particular, the straight line preserves the subalgebra $\mathfrak{osp}(4^*|4)$ of the full superconformal symmetry $\mathfrak{psu}(2,2|4)$ of $\Nm = 4$ SYM. One can then consider correlation functions of local operators in the presence of the line defect, which inherit some of the constraints from the parent theory.

There are several types of configurations one might consider in this context. One option is to study correlators involving \textit{bulk} operators in the presence of the line defect. The simplest correlators that are not fixed kinematically are two-point functions. This setup was studied at weak coupling in \cite{Buchbinder:2012vr,Barrat:2020vch} and at strong coupling in \cite{Giombi:2012ep,Buchbinder:2012vr,Liendo:2016ymz,Barrat:2021yvp}. There is also an exact topological limit that has been studied using localization \cite{Giombi:2009ds,Giombi:2012ep,Beccaria:2020ykg}.  A second option is to consider a mixture of \textit{bulk} local operators and \textit{defect} insertions along the line, although very little work has been done on this setup, except for the case of one point functions \cite{Okuyama:2006jc,Giombi:2009ds,Billo:2018oog}.

The third possibility is to focus exclusively on \textit{defect} insertions along the line, without \textit{bulk} fields. This is the configuration that we study in this paper, for which the correlators are described by a  $1d$ (non-local) CFT.  Four-point functions of these defect operators have been studied both at weak \cite{Kiryu:2018phb} and strong \cite{Giombi:2017cqn, Liendo:2018ukf,Ferrero:2021bsb} coupling. In addition, numerical results have been obtained for arbitrary coupling using a mix of integrability and bootstrap techniques \cite{Cavaglia:2021bnz}. For the simplest unprotected operator, the scaling dimension has been determined up to five loops at weak coupling \cite{Grabner:2020nis}, and to four loops at strong coupling \cite{Ferrero:2021bsb}. Finally, we note that multipoint correlators have been studied so far only in a special topological limit using supersymmetric localization \cite{Giombi:2018qox,Giombi:2018hsx}, and that the large-charge limit of two-point functions was computed recently in \cite{Giombi:2021zfb}.\footnote{Line defects have also been studied for $\Nm=2$ theories in $4d$ \cite{Gimenez-Grau:2019hez}  and for ABJM theory in \cite{Bianchi:2020hsz}.}

In this paper we compute multipoint correlation functions of local single-trace operators of the $1d$ CFT that describes the supersymmetric Wilson line.  Individual multipoint correlators contain information about an infinite number of lower-point functions by means of the OPE. This makes multipoint correlators a prime target for the conformal bootstrap program, and their study might become a powerful tool in the near future.\footnote{Recently, multipoint conformal conformal blocks have been identified as eigenfunctions of a set of Hamiltonians derived from Gaudin models \cite{Buric:2020dyz,Buric:2021ywo} (see also \cite{Rosenhaus:2018zqn}).} 
Our main goal here is to derive a formula for these $1d$ correlators at next-to-leading order in the weak coupling limit, similar to what was done for $\mathcal{N}=4$ SYM without defects in \cite{Drukker:2008pi,Drukker:2009sf}. 

The structure of the paper is as follows. In section \ref{sec:preliminaries} we introduce the basics of the $1d$ CFT and the action to be used for the subsequent computations. Section \ref{sec:correlators} is dedicated to the construction of a recursive formula at next-to-leading order for $n$-point functions of identical operators of length $\Delta = 1$. In section \ref{sec:results} we then show that this result also encodes correlators of operators with arbitrary scaling dimension, and present a few explicit examples. We end this section by conjecturing multipoint Ward identities satisfied by all the correlators we calculated. In section \ref{sec:conclusions} we conclude with a discussion of our main results and possible future directions.
\section{Preliminaries}
\label{sec:preliminaries}

In this section we introduce the $1d$ defect CFT induced by the half-BPS Wilson-line defect, focusing in particular on the structure of correlation functions of protected single-trace operators. We also state the elementary Feynman rules of the four-dimensional $\Nm = 4$ SYM that will be used in our calculations.

\subsection{$1d$ defect CFT}
\label{subsec:defectCFT}

Let us start by writing the Maldacena-Wilson line operator in $4d$ $\mathcal{N}=4$ SYM, i.e. the operator defined in \eqref{eq:WC} with the path $C$ being a straight line:
\begin{equation}
\Wl := \frac{1}{N} \tr \mathcal{P} \exp \int_{-\infty}^{\infty} d\tau\, \bigl( i 
\dot{x}^\mu(\tau) \tensor{A}{_{\smash{\mu}}}(x) + | \dot{x} (\tau)|\, \tensor{\theta}{^i} \tensor{\phi}{_i}(x) \bigr) \,,
\label{eq:wilsonline}
\end{equation}
where $\theta$ is an $SO(6)$ vector satisfying $\theta^2 = 1$. A common choice, which we also adopt in this work, is $\theta := (0,0,0,0,0,1)$.  Note that we have Wick rotated (\ref{eq:WC}) to Euclidean space and that we take the line to extend in the Euclidean time direction, i.e. $\dot{x}_\mu = (0,0,0,1)$ and $|\dot{x}|=1$. The Wilson line is a half-BPS operator and its expectation value is just
\begin{equation}
\vev{\Wl} = 1\,.
\end{equation}

This extended operator can be viewed as a \textit{defect} if it is considered to be part of the vacuum of the theory. In this case the conformal symmetry of $\Nm = 4$ SYM is broken in a controlled manner from $SO(4,2)$ to $SO(1,2) \times SO(3)$. If we restrict our attention to operators inserted \textit{on} the line, then $SO(1,2)$ corresponds to a $1d$ CFT, for which the representations carry the quantum number $\Delta$ (the scaling dimension), while the group corresponding to rotations orthogonal to the defect, $SO(3)$, refers to an \textit{internal} symmetry (spin) with quantum number $s$.

The $R$-symmetry is also broken in the presence of the defect, from $SO(6)_R$ to $SO(5)_R$. This happens because the Wilson line defined in \eqref{eq:wilsonline} only couples to a subset of scalar fields in the $R$-symmetry space, and in particular, with our choice of the polarization vector $\theta$, it turns out that $SO(5)_R$ refers to the five other scalar fields $\phi^{1, \ldots, 5}$ which do not couple to the line, while $\phi^6$ does.  Note that the full superconformal algebra $\mathfrak{psu}(2,2|4)$ of $\Nm=4$ SYM breaks in this setup into the $\Nm=8$ superconformal quantum mechanics algebra $\mathfrak{osp}(4^*|4)$.

In this work we will focus on \textit{single-trace} representations of this algebra,  in particular on the scalar sector ($s=0$). Just as in the bulk theory, we can construct protected operators out of the five scalars which do not couple to the Wilson line\footnote{In principle, \textit{multi-trace} operators with the same quantum numbers can also be constructed:
\begin{equation}
\Op_{J|\mathbf{K}}(\tau):= \Wl [(u \cdot \phi)^J(\tau)] \tr\, (u \cdot \phi)^{K_1}(\tau) \dots \tr\,(u \cdot \phi)^{K_n}(\tau) \,,
\label{eq:multitrace}
\end{equation}
where $\mathbf{K} := (K_1, \ldots, K_n)$ encodes the number of traces outside the Wilson line. These operators are half-BPS and have the protected scaling dimension $\Delta = J+ K_1 + \ldots + K_n$.}:
\begin{equation}
\Op_{\Delta} (u, \tau) :=  \Wl [ (u \cdot \phi)^{\Delta} (\tau) ]\,,
\label{eq:defectoperators}
\end{equation}
where $u$ is a complex vector satisfying $u^2 = 0$ and $u \cdot \theta = 0$, while $\Wl [ \ldots ]$ is defined as
\begin{equation}
\Wl [\mathcal{O}_1(\tau_{1}) \ldots \mathcal{O}_n(\tau_{n}) ] := \frac{1}{N} \tr \Pm \left[ \mathcal{O}_1 \ldots \mathcal{O}_n \exp \int_{-\infty}^{\infty} d\tau \bigl( i \tensor{\dot{x}}{^{\smash{\mu}}} \tensor{A}{_{\smash{\mu}}} + | \dot{x} |\, \tensor{\phi}{^6} \bigr) \right]\,,
\label{eq:wilsonlineop}
\end{equation}
where we suppressed the dependency on $\tau_1, \ldots, \tau_n$ and $u_1, \ldots, u_n$ (for the local insertions) and on $\tau$ (for the Wilson line itself) for compactness. Such operators are called \textit{insertions} on the line, since they are effectively inserted inside the trace of the Wilson line. 

\subsection{Correlation functions}
\label{subsec:correlators}

The $n$-point correlation functions of the defect single-trace operators introduced in the previous subsection are to be understood in the following way:
\begin{equation}
\vev{\Op_{\Delta_1}\, \ldots \Op_{\Delta_n}}_{1d} := \frac{1}{N} \vev{\tr \Pm \left[ (u \cdot \phi)^{\Delta_1}\, \ldots\, (u \cdot \phi)^{\Delta_n} \exp \int_{-\infty}^{\infty} d\tau \bigl( i \tensor{\dot{x}}{^{\smash{\mu}}} \tensor{A}{_{\smash{\mu}}} + | \dot{x} |\, \tensor{\phi}{^6} \bigr) \right]}_{4d}\,.
\label{eq:correlators}
\end{equation}
As indicated by the subscripts, the expectation value on the LHS refers to the correlators of the $1d$ CFT, while the one on the RHS corresponds to correlators in the $4d$ $\Nm = 4$ SYM theory. In the following, the correlators are always meant to be in the $1d$ theory and so from now on we drop these subscripts.

We emphasize here that the correlation functions described in \eqref{eq:correlators} contain only \textit{one} color trace, as opposed to the bulk theory where each operator carries its own trace. This property will be crucial for our analysis as it allows to take a limit where two adjacent operators are brought close to each other in order to give another single-trace operator with higher $R$-charge. We call this limit \textit{pinching}, which we will describe in more detail at the end of this subsection.

Because of conformal symmetry, the two-point functions are given by
\begin{equation}
\vev{\Op_{\Delta_1} (u_1, \tau_1) \Op_{\Delta_2} (u_2,\tau_2)} = n_{\Delta_1}\, \delta_{\Delta_1, \Delta_2} (12)^{\Delta_1}\,,
\end{equation}
where $(ij)$ encodes the propagator of the scalars. In our case, because of $R$-symmetry,  this propagator also contains the $R$-symmetry vectors in the following way
\begin{equation}
(ij) := \frac{(u_i \cdot u_j)}{\tau_{ij}^2}\,,
\label{eq:ij}
\end{equation}
with $\tau_{ij} := \tau_i - \tau_j$. Note that the scaling dimension of the fundamental scalar field in this $1d$ CFT is $\Delta = 1$ due to its origin from a $4d$ bulk theory. The normalization constants $n_\Delta$ are known to be non-trivial functions of the coupling and can be computed from localization or from a Feynman diagrammatic expansion \cite{Giombi:2018qox,Giombi:2018hsx}.

Three-point functions are also kinematically fixed by conformal symmetry and read
\begin{equation}
\vev{\Op_{\Delta_1} (u_1,\tau_1) \Op_{\Delta_2} (u_2,\tau_2) \Op_{\Delta_3} (u_3,\tau_3)} = \lambda_{123}\, (12)^{\Delta_{123}} (23)^{\Delta_{231}} (31)^{\Delta_{312}}\,,
\end{equation}
with $\Delta_{ijk} := \Delta_i + \Delta_j - \Delta_k$. Note that since the scaling dimensions of the operators are protected, only the OPE coefficients $\lambda_{ijk}$ can receive quantum corrections here.

For higher $n$-point functions, conformal symmetry is not strong enough to fix the kinematical form of the correlators. It is convenient to consider the following factorized form of the multipoint correlation functions:
\begin{eqnarray}
\vev{\Op_{\Delta_1} \ldots \Op_{\Delta_n}} = \Km_{\Delta_1 \ldots \Delta_n} 
\, \Am_{\Delta_1 \ldots \Delta_n} (\chi_i\,, r_i\,, s_i\,, t_{ij})\,,
\label{eq:correlatorsdef}
\end{eqnarray}
where $\chi_i$ are the spacetime cross-ratios, $r_i\,, s_i\,, t_{ij}$ are the $R$-symmetry cross-ratios, all to be defined shortly, and $\Km_{\Delta_1 \ldots \Delta_n}$ corresponds to a (super)conformal prefactor, chosen such that the reduced correlator $\Am_{\Delta_1 \ldots \Delta_n}$ depends only on these cross-ratios. Note that we always choose the most convenient $\Km_{\Delta_1 \ldots \Delta_n}$ that we can think of for each correlator, instead of sticking to a general definition. In $1d$ there are only $n-3$ spacetime cross-ratios and $n(n-3)/2$ $R$-symmetry cross-ratios, which we are going to carefully define in the following.

First of all we choose the spacetime cross-ratios such that the following limit holds:
\begin{equation}
(\chi_1, \chi_2, \ldots, \chi_{n-3}) \overset{(\tau_1, \tau_{n-1}, \tau_n) \rightarrow (0,1,\infty)}{\longrightarrow} (\tau_2, \ldots, \tau_{n-2})\,.
\end{equation}
This results in the following expressions:
\begin{equation}
\chi_1 = \frac{\tau_{12} \tau_{(n-1)n}}{\tau_{1(n-1)} \tau_{2n}}\,, \quad \chi_2 = \frac{\tau_{13} \tau_{(n-1)n}}{\tau_{1(n-1)} \tau_{3n}}\,, \ldots,
\quad \chi_i = \frac{\tau_{1(i+1)} \tau_{(n-1)n}}{\tau_{1(n-1)} \tau_{(i+1)n}}\, .
\end{equation}
Notice that the $1-\chi_i$ (which would be independent cross-ratios in a higher-dimensional CFT) are given by
\begin{equation}
1-\chi_1 = \frac{\tau_{1n} \tau_{2(n-1)}}{\tau_{1(n-1)} \tau_{2n}}\,, \quad 1-\chi_2 = \frac{\tau_{1n} \tau_{3(n-1)}}{\tau_{1(n-1)} \tau_{3n}}\,, \ldots,
\quad 1-\chi_i = \frac{\tau_{1n} \tau_{(i+1)(n-1)}}{\tau_{1(n-1)} \tau_{(i+1)n}}\, .
\end{equation}

For the $R$-symmetry cross-ratios, we start by defining the $r_i$ such that the indices have a one-to-one correspondence with the spacetime cross-ratios, i.e.
\begin{equation}
r_1 = \frac{(u_1 \cdot u_2) (u_{n-1} \cdot u_n)}{(u_1 \cdot u_{n-1}) (u_2 \cdot u_n)}\,, r_2 = \frac{(u_1 \cdot u_3) (u_{n-1} \cdot u_n)}{(u_1 \cdot u_{n-1}) (u_3 \cdot u_n)}\,, \ldots \, ,
r_i = \frac{(u_1 \cdot u_{i+1}) (u_{n-1} \cdot u_n)}{(u_1 \cdot u_{n-1}) (u_{i+1} \cdot u_n)}\, .
\end{equation}
This correspondence implies that we have $n-3$ $r_i$.

Then the $s_i$ are defined in analogy to $1-\chi_i$. This gives the following $n-3$ cross-ratios:
\begin{equation}
s_1 = \frac{(u_1 \cdot u_n) (u_2 \cdot u_{n-1})}{(u_1 \cdot u_{n-1}) (u_2 \cdot u_n)}\,, s_2 = \frac{(u_1 \cdot u_n) (u_3 \cdot u_{n-1})}{(u_1 \cdot u_{n-1}) (u_3 \cdot u_n)}\,, \ldots\, , 
s_i = \frac{(u_1 \cdot u_n) (u_{i+1} \cdot u_{n-1})}{(u_1 \cdot u_{n-1}) (u_{i+1} \cdot u_n)}\,.
\end{equation}

It is well-known that the correlators defined in \eqref{eq:correlatorsdef} are \textit{topological} when we set $u_i \cdot u_j = \tau_{ij}^2$, i.e.~the functions $\Am_{\Delta_1 \ldots \Delta_n}$ are constant in this limit, in the sense that they do not depend on the variables $u$ and $\tau$ \cite{Drukker:2009sf}\footnote{They can however still depend non-trivially on the 't Hooft coupling $\lambda$ and the number of colors $N$.}. Therefore, we find it useful to define the remaining $(n-3)(n-4)/2$ $R$-symmetry cross-ratios $t_{ij}$ in a way such that they reduce in the topological sector to the analogue of the following spacetime cross-ratios:
\begin{equation}
t_{ij} \to (\chi_i - \chi_j)^2\,,
\end{equation}
namely
\begin{equation}
t_{ij}= \frac{(u_{i+1} \cdot u_{j+1})(u_1 \cdot u_n)(u_{n-1} \cdot u_n)}{(u_1 \cdot u_{n-1})(u_{i+1} \cdot u_n)(u_{j+1} \cdot u_n)}
\end{equation}
with $i=[1,n-4]\, , j=[i+1, n-3]$ and $i<j$.

As mentioned before, since the operators are inserted inside the trace of the Wilson line, the \textit{pinching} of two operators or more produces single-trace operators again, e.g.
\begin{equation}
\vev{\Op_1 (u_1, \tau_1) \ldots \Op_1 (u_{n-1},\tau_{n-1}) \Op_1 (u_n,\tau_n)} \overset{n \to n-1}{\rightarrow} \vev{\Op_1 (u_1, \tau_1) \ldots  \Op_2 (u_{n-1},\tau_{n-1})}\,.
\end{equation}
This pinching technique will be used in section \ref{sec:results} to construct arbitrary correlators from $\vev{\Op_1 \ldots \Op_1}$. 
The fact that correlators of single-trace operators close under pinching has an interesting consequence: the \textit{information} needed to solve the scalar sector in this theory is very much reduced compared e.g. to its bulk counterpart!

\subsection{Bulk action and propagators}
\label{subsec:action}

As stated above, although the correlators satisfy the axioms of a $1d$ CFT, we perform the computations using the $4d$ action of $\Nm = 4$ SYM. The latter is given by (including ghosts and gauge fixing)
\begin{align}
S &= \frac{1}{g^2} \int d^4 x\ \text{Tr} \left\lbrace \frac{1}{2} \tensor{F}{_{\mu\nu}} \tensor{F}{^{\mu\nu}} + \tensor{D}{_\mu} \tensor{\phi}{_i} \tensor{D}{^\mu} \tensor{\phi}{^i} - \frac{1}{2} [ \tensor{\phi}{_i} , \tensor{\phi}{_j} ] [ \tensor{\phi}{^i} , \tensor{\phi}{^j} ] \right. \notag \\
& \left. \qquad \qquad \qquad \qquad \qquad \qquad + i \bar{\psi} \tensor{\gamma}{^\mu} \tensor{D}{_\mu} \psi + \bar{\psi} \tensor{\Gamma}{^i} [ \tensor{\phi}{_i} , \psi ] + \tensor{\partial}{_\mu} \bar{c} \tensor{D}{^\mu} c + \xi \left( \tensor{\partial}{_\mu} \tensor{A}{^\mu} \right)^2 \right\rbrace\;.
\label{eq:action}
\end{align}
Our conventions are collected in appendix \ref{sec:insertion}. The resulting propagators in Feynman gauge ($\xi = 1$) take the following form in configuration space:
\begin{subequations}
\begin{align}
\text{Scalars:} \qquad 
& \propagatorS = g^2 \tensor{\delta}{_{ij}} \tensor{\delta}{^{ab}} I_{12}\;, \label{subeq:propagatorS} \\
\text{Gluons:} \qquad 
& \propagatorG = g^2 \tensor{\delta}{_{\mu\nu}} \tensor{\delta}{^{ab}} I_{12}\;, \label{subeq:propagatorG} \\
\text{Gluinos:} \qquad 
& \propagatorF = i g^2 \tensor{\delta}{^{ab}} \slashed{\partial}_{\Delta} I_{12}\;, \label{subeq:propagatorF} \\
\text{Ghosts:} \qquad 
& \propagatorGh = g^2 \tensor{\delta}{^{ab}} I_{12}\;, \label{subeq:propagatorGh}
\end{align}
\label{eq:propagators}%
\end{subequations}
where we have defined for brevity
\begin{equation}
I_{ij} := \frac{1}{(2\pi)^2 x_{ij}^2}\;,
\label{eq:I12}
\end{equation}
with $x^{\mu}_{ij} := x^{\mu}_i - x^{\mu}_j$ and
\begin{equation*}
\slashed{\partial}_{\Delta} := \gamma \cdot \frac{\partial}{\partial \Delta}\;, \qquad \qquad \Delta^{\mu} := x_{12}^{\mu}\;,
\end{equation*}
with $\gamma_\mu$ the Dirac matrices. The Feynman rules are easy to obtain, and a set of convenient insertion rules can be found in appendix \ref{sec:insertion}.
\section{Correlation functions of $\Delta = 1$ operators}
\label{sec:correlators}

We now study the most elementary class of operators, which consist of a single scalar field insertion on the Wilson line.   We restrict ourselves to the large $N$ limit, and use the 't Hooft coupling $\lambda := g^2 N$ as the parameter of the perturbative expansion.

For compactness, we define the following shorthand notation:
\begin{equation} 
A_n(1,\ldots , n) := \vev{\Op_1 (u_1,\tau_1) \ldots \Op_1 (u_n, \tau_n)}\,,
\end{equation}
which we shall be studying at leading order (LO) and next-to-leading (NLO) order precision.  Notice that for odd $n$ the correlators $A_n$ vanish due to $R$-symmetry.

\subsection{Leading order}
\label{subsec:tree}

We start by deriving a leading-order formula. For operators of scaling dimension $\Delta = 1$, it is easy to find a recursive expression for $n$-point functions at leading order.  In fact this problem is related to a more mathematical one concerning meanders and arch statistics, which was already solved in \cite{DiFrancesco:1995cb}. Adapting (3.1) of that paper to our case of interest, we obtain the recursion
\begin{equation}
A_n^\text{\tiny{LO}}(1,\ldots , n)  = \sum_{j=0}^{\frac{n}{2}-1} A_2^\text{\tiny{LO}}(1,2j+2) A_{2j}^\text{\tiny{LO}}(2,\ldots, 2j+1) A_{n-2-2j}^\text{\tiny{LO}}(2j+3,\ldots, n)\,,
\label{eq:recursiontree}
\end{equation}
which can be represented diagrammatically as
\begin{equation}
A_n^\text{\tiny{LO}}(1,\ldots , n) = \sum_{j=0}^{\frac{n}{2}-1} \recursiontree\,,
\end{equation}
where \treelevelblob \, stands for the leading-order correlation function of the appropriate lengths.

In the expression above, the starting values for the recursion are given by the vacuum expectation value and by the two-point functions:
\begin{equation} 
A_0^\text{\tiny{LO}} = 1\,, \qquad A_2^\text{\tiny{LO}}(i,j) = \frac{\lambda}{8 \pi^2} (ij)\,,
\label{eq:startingvalstree}
\end{equation}
with $(ij)$ defined in \eqref{eq:ij}. As mentioned above, only correlators with an even number of operators are non-vanishing, and two- and four-point functions can be compared with the results of \cite{Kiryu:2018phb}, with which they agree perfectly.

\subsection{Next-to-leading order}
\label{subsec:oneloop}

At next-to-leading order the situation becomes more intricate, not only because of the appearance of $4d$ vertices, but also because some of them couple to the Wilson line. Nevertheless, we can write a recursive diagrammatic formula, which produces all the relevant Feynman diagrams for an arbitrary $n$-point function of $\Op_1$ operators:
\begin{align}
A_n^\text{\tiny{NLO}}(1,\ldots , n) =& \sum_{i=1}^{n-3} \sum_{j=i+1}^{n-2} \sum_{k=j+1}^{n-1} \sum_{l=k+1}^{n} \recursionFour \notag \\
& + \sum_{i=1}^{n-1} \sum_{j=i+1}^{n} \Biggl( \recursionSE + \sum_{k=0}^{i-1} \recursionYNone  \notag \\
& \qquad\quad + \sum_{k=i}^{j-1} \recursionYNtwo + \sum_{k=j}^{n+1} \recursionYNthree \Biggr)  \notag \\
&  + \sum_{i=1}^{n-3} \sum_{j=i+3}^{n} \recursionbridge\,.
\label{eq:recursionloop}
\end{align}

Let us analyze this expression. First of all,  its recursiveness is encoded in two different aspects: on one hand in the dependency on the leading-order correlators, which obey the relation given in \eqref{eq:recursiontree},  on the other hand on the NLO correlators, as one can note from the last term of the sum, where one should insert the NLO correlation function with $n=j-i-1$\footnote{Note that it is recursive with respect to correlators on a \textit{finite} line and not extending to $\pm \infty$. This is explained in more detail below \eqref{eq:formulaOneloop}.}. 

Let us now describe in detail each term that appears in \eqref{eq:recursionloop}. The first one corresponds to the possible NLO insertions involving four scalar lines, namely
\begin{equation}
\vertexFourno\ := \vertexSSSSno\ + \Hinsertoneno\ + \Hinserttwono\ ,
\end{equation}
and the associated diagram reads
\begin{align}
\recursionFour &= A_4^4 (i,j,k,l) A_{i-1}^\text{\tiny{LO}} (1, \ldots, i-1) A_{j-i-1}^\text{\tiny{LO}} (i+1, \ldots, j-1) \notag \\
& \qquad\qquad \raisebox{2ex}{$\times A_{k-j-1}^\text{\tiny{LO}} (j+1, \ldots, k-1) A_{l-k-1}^\text{\tiny{LO}} (k+1, \ldots, l-1)$} \notag \\
& \qquad\qquad\qquad \times A_{n-l}^\text{\tiny{LO}} (l+1, \ldots, n)\,.
\label{eq:contribX}
\end{align}
Using the insertion rules \eqref{eq:vertexSSSS} and \eqref{eq:Hinsertone}, we define the following four-point correlator as a NLO building block:
\begin{align}
A_4^4 (i,j,k,l) :=& \frac{\lambda^3}{8} \left[\left(2 (u_i \cdot u_k) (u_j \cdot u_l) -(u_i \cdot u_l)(u_j \cdot u_k)-(u_i \cdot u_j)(u_k \cdot u_l)\right) X_{ijkl} \right.  \notag \\
& \left. + (u_i \cdot u_l)(u_j \cdot u_k) I_{il} I_{jk} F_{il;jk}-(u_i \cdot u_j)(u_k \cdot u_l) I_{ij} I_{kl} F_{ij;kl} \right]\,.
\end{align}
The results for the integrals $X_{ijkl}$ and $F_{ij;kl}$ associated to the insertion rules can be found in appendix \ref{subsec:standardintegrals}. Note that all the diagrams encompassed by this term are perfectly finite as long as the external points are distinct. 

The second and third lines in \eqref{eq:recursionloop}, corresponding to self-energy and $Y$-diagrams, also add up to a finite result since divergences coming from the self-energy cancel with the ones arising in the $Y$-diagrams.  To be precise, the self-energy diagrams read
\begin{equation}
\recursionSE = A_2^\text{\tiny{SE}} (i,j) A_{i-1}^\text{\tiny{LO}} (1, \ldots, i-1) A_{j-i-1}^\text{\tiny{LO}} (i+1,  \ldots, j-1) A_{n-j}^\text{\tiny{LO}} (j+1, \ldots, n)\,,
\label{eq:formulaSE}
\end{equation}
where we use as a building block the well-known expression for the scalar propagator at NLO (see \eqref{eq:selfenergy})
\begin{equation}
A_2^\text{\tiny{SE}} (i,j) := - \lambda^2\, (u_i \cdot u_j)\, Y_{iij}\,.
\label{eq:A2SE}
\end{equation}
We turn now our attention to the $Y$-diagrams. Note that inserting a gluon field on the Wilson line corresponds to expanding the exponential of \eqref{eq:correlators} up to first order, and thus it results in a one-dimensional integral between the points before and after the insertion. For example,
\begin{equation}
\recursionYexplainone := \int_{\tau_k}^{\tau_{k+1}} d\tau_{\alpha}\ \recursionYexplaintwo\,.
\end{equation}

In order to show that the divergences cancel with the ones coming from the self-energy graphs, it is convenient to express the $Y$-diagrams in a different way:
\begin{align}
\sum_{k=0}^{i-1} \recursionYNone =&\ \recursionYLidone \notag \\
& - \sum_{\alpha = 1}^{i-2} \sum_{\beta = \alpha+1}^{i-1} \recursionYLtwo\,,
\label{eq:reformulateY}
\end{align}
where the red dots indicate the places where the gluon line should be connected. The sum over $i, j$ is implied here. 
It should be clarified that the diagrams on the right-hand side should \textit{not} be considered non-planar when the gluon line is crossing a scalar line. Similarly, such a crossing does \textit{not} generate an additional $4$d vertex. Here the dots are intended to only indicate the range of integration.
It is easy to check that the relation \eqref{eq:reformulateY} holds by rewriting the integration limits of the left-hand side as $\int_a^b = \int_{-\infty}^{\infty} - \int_{-\infty}^a - \int_b^{\infty}$.

The same can be performed for the other two terms with $Y$-vertices, and we are left with the following diagrams to compute:
\begin{gather*}
\recursionYone - \sum_{\alpha = 1}^{i-2} \sum_{\beta = \alpha+1}^{i-1} \recursionYLtwo \\ - \sum_{\alpha = i+1}^{j-2} \sum_{\beta = \alpha+1}^{j-1} \recursionYCtwodue -  \sum_{\alpha = i+1}^{n-1} \sum_{\beta = \alpha+1}^{n} \recursionYRtwo\,,
\end{gather*}
which of course also have to be summed over $i$ and $j$. By doing so, we have isolated the divergences inside the first term, since the integration ranges of the other terms do not include the points $\tau_i$ and $\tau_j$. Moreover, since the limits of integration of the first term are $-\infty$ and $+\infty$, we can perform the integral analytically and extract the divergences. We then find
\begin{equation}
\recursionYone = A_2^\text{\tiny{Y,div}} (i,j) A_{i-1}^\text{\tiny{LO}} (1, \ldots, i-1) A_{j-i-1}^\text{\tiny{LO}} (i+1,  \ldots, j-1) A_{n-j}^\text{\tiny{LO}} (j+1, \ldots, n)\,,
\label{eq:formulaYdiv}
\end{equation}
with
\begin{equation}
A_2^\text{\tiny{Y,div}} (i,j) := \lambda^2 (u_i \cdot u_j)\, Y_{iij} + \frac{\lambda^2}{4} (u_i \cdot u_j)\, T_{ij;0(n+1)}\,,
\label{eq:A2Ydiv}
\end{equation}
where the integral $T_{ij;kl}$ is defined in \eqref{eq:Tdef}, with the subscripts $0$ and $n+1$ referring respectively to $-\infty$ and $+\infty$. The case $(0,n+1) = (-\infty,+\infty)$ can be found in \eqref{eq:T1}. Since the recursive structures of \eqref{eq:formulaSE} and \eqref{eq:formulaYdiv} are identical, it is clear that the divergences of \eqref{eq:A2SE} and \eqref{eq:A2Ydiv} cancel perfectly, and since the remaining $T$-integrals are finite, we are left with a finite expression.

The remaining term on the RHS of \eqref{eq:reformulateY} is also finite and reads
\begin{align}
\recursionYLtwo =& A_4^\text{\tiny{Y}} (i,j,\alpha,\beta) A^\text{\tiny{LO}}_{\alpha-1} (1, \ldots, \alpha-1) \notag \\
& \raisebox{2ex}{$\times A^\text{\tiny{LO}}_{\beta-\alpha-1} (\alpha+1, \ldots, \beta-1) A^\text{\tiny{LO}}_{i-\beta-1} (\beta+1, \ldots, i-1)$} \notag \\
& \times A^\text{\tiny{LO}}_{j-i-1} (i+1, \ldots, j-1) A^\text{\tiny{LO}}_{n-j} (j+1, \ldots, n)\,,
\label{eq:Ynodiv1}
\end{align}
where the starting point evaluates to
\begin{equation}
A_4^\text{\tiny{Y}} (i,j,\alpha,\beta) := \frac{\lambda^3}{32 \pi^2 \tau_{\alpha\beta}^2} (u_\alpha \cdot u_\beta) (u_i \cdot u_j)\, T_{ij;\alpha\beta}\,.
\label{eq:A4Y}
\end{equation}

The two other terms (center and right) can be implemented in the very same way. For the center term we have
\begin{align}
\recursionYCtwodue =& A_4^\text{\tiny{Y}} (i,j,\alpha,\beta) A^\text{\tiny{LO}}_{i-1} (1, \ldots, i-1) \notag \\
& \raisebox{2ex}{$\times A^\text{\tiny{LO}}_{\alpha-i-1} (i+1, \ldots, \alpha-1) A^\text{\tiny{LO}}_{\beta-\alpha-1} (\alpha+1, \ldots, \beta-1)$} \notag \\
& \times A^\text{\tiny{LO}}_{j-\beta-1} (\beta+1, \ldots, j-1) A^\text{\tiny{LO}}_{n-j} (j+1, \ldots, n)\,,
\label{eq:Ynodiv2}
\end{align}
while for the right one we obtain
\begin{align}
\recursionYRtwo =& A_4^\text{\tiny{Y}} (i,j,\alpha,\beta) A^\text{\tiny{LO}}_{i-1} (1, \ldots, i-1) \notag \\
& \raisebox{2ex}{$\times A^\text{\tiny{LO}}_{j-i-1} (i+1, \ldots, j-1) A^\text{\tiny{LO}}_{\alpha-j-1} (j+1, \ldots, \alpha-1)$} \notag \\
& \times A^\text{\tiny{LO}}_{\beta-\alpha-1} (\alpha+1, \ldots, \beta-1) A^\text{\tiny{LO}}_{n-\beta} (\beta+1, \ldots, n)\,.
\label{eq:Ynodiv3}
\end{align}
The integrals in \eqref{eq:A4Y} give different results depending on the ordering of the variables $\tau_i, \tau_j, \tau_\alpha, \tau_\beta$. The results have been collected in \eqref{eq:Tint}.

The last term of the formula given in \eqref{eq:recursionloop} is recursive with respect to the full next-to-leading order formula:
\begin{equation}
\recursionbridge = A_{j-i-1}^\text{\tiny{NLO}} (i+1,\ldots,j-1) A_{i-1}^\text{\tiny{LO}} (1,\ldots,i-1) A_2^\text{\tiny{LO}}(i,j) A_{n-j}^\text{\tiny{LO}} (j+1,\ldots,n)\,.
\label{eq:formulaOneloop}
\end{equation}
However, we must be careful here, because the limits of integration for the inserted NLO expression are \textit{not} $\pm \infty$ but $(0, n+1) = (i,j)$ for the $Y$-diagrams in \eqref{eq:A2Ydiv}.

The recursion relation \eqref{eq:recursionloop} and the associated expressions are enough to determine $n$-point functions of protected operators of length $\Delta=1$. The expressions are quite lengthy, so we will not give the results here. We refer the reader to the ancillary \textsc{Mathematica} notebook for the four- and six-point functions, while for the rest of this paper we focus on correlators involving operators of arbitrary scaling dimensions.
\section{Correlation functions of arbitrary operators}
\label{sec:results}
\begingroup
\allowdisplaybreaks

We will now present correlators with operators of arbitrary scaling dimensions, using the formulae presented in the previous section and the pinching technique mentioned in section \ref{subsec:correlators}. In this way we can derive arbitrary $n$-point functions of protected operators, for which we give some low-lying examples. Surprisingly, we observe that all our correlators satisfy certain differential equations. We conjecture that these equations also hold non-perturbatively and that they correspond to the multipoint extension of the superconformal Ward identities, currently known for the case $n=4$ only \cite{Liendo:2018ukf}.

\subsection{Extremal correlators}
\label{subsec:extremal}

We start our analysis by studying the simplest correlators. We call \textit{extremal} the correlators for which the length of one operator is equal to the sum of the lengths of all the other operators, i.e.
\begin{equation*}
\vev{\Op_{\Delta_1} (\tau_1) \ldots \Op_{\Delta_{n-1}} (\tau_{n-1}) \Op_{\Delta_n} (\tau_n)}\,,
\end{equation*}
with $\Delta_n = \Delta_1 + \ldots + \Delta_{n-1}$. It was shown in \cite{Bianchi:1999ie} that such correlators do not renormalize in the bulk, and while the next-to-leading order result is not zero in our case\footnote{We note that this result differs from equation (84) in \cite{Kiryu:2018phb}, which we believe is not applicable to the extremal case.}, we observe that the kinematics are trivial as well. Using the recursion relation, we were able to find a closed form for the extremal correlators up to next-to-leading order:
\begin{equation}
\vev{\Op_{\Delta_1} (\tau_1) \ldots \Op_{\Delta_n} (\tau_n)} \raisebox{-.5ex}{$\bigr|$}_{\Delta_n = \Delta_1 + \ldots + \Delta_{n-1}} = \frac{\lambda^{\Delta_n}}{2^{3 \Delta_n} \pi^{2\Delta_n}} \left( 1 - \frac{\lambda}{24} + \Op(\lambda^2) \right) \prod_{j=1}^{n-1} (jn)^{\Delta_j}\,,
\end{equation}
where $(jn)$ corresponds to the free propagator defined in \eqref{eq:ij}. Remarkably all the $X$, $F$ and $T$-integrals containing transcendental functions cancel each other, and the product on the right-hand side can be interpreted as the superconformal prefactor $\Km$,  that we introduced in (\ref{eq:correlatorsdef}). The result can thus simply be reformulated as
\begin{equation}
\Am_{\Delta_1 \ldots \Delta_n} \raisebox{-.5ex}{$\bigr|$}_{\Delta_n = \Delta_1 + \ldots + \Delta_{n-1}} = \frac{\lambda^{\Delta_n}}{2^{3 \Delta_n} \pi^{2\Delta_n}} \left( 1 - \frac{\lambda}{24} + \Op(\lambda^2) \right)\,.
\end{equation}
This expression can be checked against the localization results of \cite{Giombi:2018qox}, with which it agrees perfectly.

\subsection{Two-, three- and four-point functions}
\label{subsec:checks}

The formulae presented in section \ref{sec:correlators} can also be checked against already existing results in the literature. One can easily obtain the closed form for two-point functions by taking correlators with an even number of $\Op_1$ and pinching each half together. This results in
\begin{equation}
n_\Delta = \frac{\lambda^\Delta}{2^{3\Delta} \pi^{2\Delta}} \left( 1 - \frac{\lambda}{24} + \Op(\lambda^2) \right)\,,
\label{eq:twopt}
\end{equation}
in perfect agreement with \cite{Kiryu:2018phb}.

Three-point functions can be obtained similarly, and we find
\begin{equation}
\lambda_{123} = \left( \frac{\sqrt{\lambda}}{2\sqrt{2} \pi} \right)^{\Delta_1+\Delta_2+\Delta_3} \left( 1 - \frac{\lambda}{24} (\delta_{\Delta_1,\Delta_2+\Delta_3} + \delta_{\Delta_2,\Delta_3+\Delta_1} + \delta_{\Delta_3,\Delta_1+\Delta_2}) \right)\,,
\label{eq:threept}
\end{equation}
which matches again \cite{Kiryu:2018phb}. For all the cases we looked at, we observed perfect agreement between (84) in \cite{Kiryu:2018phb} and the pinching of our NLO formula. 

We wish now to illustrate how the recursion relations given in \eqref{eq:recursiontree} and \eqref{eq:recursionloop} work for a simple example, i.e. the four-point function of operators $\Op_1$. Using the notation introduced in section \ref{subsec:correlators}, the reduced correlator can be extracted from the full correlator following
\begin{equation}
\vev{\Op_1 \Op_1 \Op_1 \Op_1} = \Km_{1111} \Am_{1111} (\chi; r,s)\,,
\end{equation}
where we choose the conformal prefactor to be
\begin{equation}
\Km_{1111} := \frac{(u_1 \cdot u_2)(u_3 \cdot u_4)}{\tau_{12}^2 \tau_{34}^2}\,,
\end{equation}
while the cross-ratios are defined as
\begin{equation}
\chi := \frac{\tau_{12} \tau_{34}}{\tau_{13} \tau_{24}}\,, \quad r := \frac{(u_1 \cdot u_2)(u_3 \cdot u_4)}{(u_1 \cdot u_3)(u_2 \cdot u_4)}\,, \quad s := \frac{(u_1 \cdot u_4)(u_2 \cdot u_3)}{(u_1 \cdot u_3)(u_2 \cdot u_4)}\,.
\end{equation}
The reduced correlator can be expanded into three $R$-symmetry channels:
\begin{equation}
\Am_{1111} := F_0 (\chi) + \frac{\chi^2}{r} F_1 (\chi) + \frac{s}{r} \frac{\chi^2}{(1-\chi)^2} F_2 (\chi)\,,
\label{eq:splittingR}
\end{equation}
where the prefactors have been chosen such that they satisfy on their own the superconformal Ward identities, which we discuss in detail later.
These channels have the following perturbative expansion:
\begin{equation}
F_j = \lambda^2 \sum_{k=0}^{\infty} \lambda^{k} F^{(k)}_j\,,
\end{equation}
and in the following we will use the recursion relations \eqref{eq:recursiontree} and \eqref{eq:recursionloop} in order to derive the two leading terms for each channel.

At leading order equation,  \eqref{eq:recursiontree} becomes
\begin{align}
\vev{\Op_1 \Op_1 \Op_1 \Op_1}_\text{LO} = & A_2^\text{LO} (1,2) A_2^\text{LO} (3,4) + A_2^\text{LO} (1,4) A_2^\text{LO} (2,3)\,,
\end{align}
which can be represented diagrammatically as
\begin{equation}
\vev{\Op_1 \Op_1 \Op_1 \Op_1}_\text{LO} =\, \fourpttreeone\, + \fourpttreetwo\,.
\end{equation}
Using the starting values given in \eqref{eq:startingvalstree} and the decomposition into $R$-symmetry channels of \eqref{eq:splittingR}, it is straightforward to obtain the following result:
\begin{subequations}
\begin{align}
F_0^{(0)} (\chi) = \frac{1}{64 \pi^4} \,, \quad F_1^{(0)} (\chi) = 0\,, \quad F_2^{(0)} (\chi) = \frac{1}{64 \pi^4}\,.
\end{align}
\end{subequations}

At next-to-leading order the recursion relation given in \eqref{eq:recursionloop} produces the following diagrams:
\begin{align}
\vev{\Op_1 \Op_1 \Op_1 \Op_1}_\text{NLO} = & \phantom{+}\,\,\, \,\fourptX\, +\, \fourptHone\, +\, \fourptHtwo \notag \\
& +\, \fourptSEone\, +\, \fourptSEtwo\, +\, \fourptSEthree\, \notag \\
& +\, \fourptYone\,+\, \fourptYtwo\, +\, \fourptYthree \notag \\
& +\, \fourptSEfour\, +\, \fourptYfour\,.
\end{align}
The first line is produced by the term \eqref{eq:contribX}, while the self-energy diagrams in the second line are generated by \eqref{eq:formulaSE} and the $Y$-diagrams from the third line come from \eqref{eq:formulaYdiv}, \eqref{eq:Ynodiv1}, \eqref{eq:Ynodiv2} and \eqref{eq:Ynodiv3}. Finally, the last line is obtained from the recursive term \eqref{eq:formulaOneloop}. Using the integrals of appendix \ref{sec:integrals}, we find the following expressions for the $R$-symmetry channels at NLO:
\begin{subequations}
\begin{align}
F_0^{(1)} (\chi) =& \frac{1}{512 \pi^6} \left(2 L_R(\chi) + \frac{\ell (\chi,1)}{1-\chi} - \frac{2\pi^2}{3} \right) \,, \\
F_1^{(1)} (\chi) =& - \frac{1}{512 \pi^6} \frac{\ell (\chi,1)}{\chi (1-\chi)}\,, \\
F_2^{(1)} (\chi) =& -\frac{1}{512 \pi^6} \left(2 L_R(\chi) - \frac{\ell (\chi,1)}{\chi} + \frac{\pi^2}{3} \right) \,,
\end{align}
\end{subequations}
which agrees with \cite{Kiryu:2018phb}.  The Rogers dilogarithm $L_R$ is defined as
\begin{equation}
L_R(\chi) := \text{Li}_2 (\chi) + \frac{1}{2} \log (\chi) \log(1-\chi)\,,
\label{eq:Rogers}
\end{equation}
and satisfying the following properties:
\begin{align}
&L_R (x) + L_R (1-x) = \frac{\pi^2}{6}\,, \\
L_R (x) + L_R(y) &= L_R (xy) + L_R \left( \frac{x(1-y)}{1-xy} \right) + L_R \left( \frac{y(1-x)}{1-xy} \right)\,.
\end{align}
We also used the following two-variable help function:
\begin{equation}
\ell(\chi_1, \chi_2) := \chi_1 \log \chi_1 - \chi_2 \log \chi_2 + (\chi_2 - \chi_1) \log (\chi_2 - \chi_1)\,.
\label{eq:smalll}
\end{equation} 
Note that $\ell(\chi,1)$ is manifestly crossing-symmetric, i.e.
\begin{equation}
\ell(\chi,1) = \ell(1-\chi,1)\,.
\end{equation}
Note also that the function $\ell$ also satisfies the following identities:
\begin{align}
&\ell(\chi_1, \chi_2) + \ell(\chi_2, \chi_1) = i \pi (\chi_1 - \chi_2)\,, \\
&\ell(\chi_1, \chi_2) = \chi_1 \chi_2\, \ell(\chi_2^{-1}, \chi_1^{-1}) \quad \text{for } 0 < \chi_1 < \chi_2 < 1\,, \\
&\ell(\chi_1, \chi_2) + \ell(1-\chi_2, 1-\chi_1) = \ell(\chi_1,1) - \ell(\chi_2,1)\,.
\end{align}

It is easy to check that the pinching of $\vev{\Op_1 \Op_1 \Op_1 \Op_1}$ into $\vev{\Op_2 \Op_2}$ matches the result given in equation \eqref{eq:twopt} for the case $\Delta = 2$. This can be obtained by taking the following limit:
\begin{equation}
(u_2, \tau_2) \to (u_1, \tau_1)\,, \quad (u_4, \tau_4) \to (u_3, \tau_3)\,.
\end{equation}

As briefly mentioned, when expressed in terms of spacetime and $R$-symmetry cross-ratios, this correlator and more generally four-point functions of arbitrary half-BPS operators satisfy the following elegant Ward identity \cite{Liendo:2016ymz}:
\begin{equation}
\left( \frac{1}{2} \partial_{\chi} + \alpha \partial_r - (1-\alpha) \partial_s \right) \,  \Am_{\Delta_1 \Delta_2 \Delta_3 \Delta_4} \raisebox{-1ex}{$\biggr |$}_{\raisebox{.75ex}{$\begin{subarray}{l} r = \alpha \chi\\
s = (1-\alpha)(1-\chi) \end{subarray}$}}=0\,,
\label{eq:WI4}
\end{equation}
which is valid for any $\alpha$ real. This differential equation encodes the constraints of superconformal symmetry on the correlators and turned out to be essential for bootstrapping the four-point function $\vev{\Op_1 \Op_1 \Op_1 \Op_1}$ at strong coupling \cite{Liendo:2018ukf,Ferrero:2021bsb}. We note that in the literature this Ward identity is given in terms of $R$-symmetry ratios which are defined slightly differently:
\begin{equation}
r = \zeta_1 \zeta_2\,, \qquad s = (1-\zeta_1)(1-\zeta_2)\,,
\end{equation}
and in that notation, \eqref{eq:WI4} splits into two independent equations:
\begin{equation}
\left. \left( \frac{1}{2} \partial_\chi + \partial_{\zeta_1} \right) \Am_{\Delta_1 \Delta_2 \Delta_3 \Delta_4} \right|_{\zeta_1 = \chi} = 0\,, \qquad \left.  \left( \frac{1}{2} \partial_\chi + \partial_{\zeta_2} \right) \Am_{\Delta_1 \Delta_2 \Delta_3 \Delta_4} \right|_{\zeta_2 = \chi} = 0\,.
\end{equation}
These formulations are equivalent and impose powerful constraints on the correlators. In the case where $\Delta_i = 1$ for all the operators, they imply remarkably that only one function $f(\chi)$ should be known in addition to the localization result in order to fix the full correlator \cite{Liendo:2018ukf}. In section \ref{subsec:conjectureWI} we conjecture a multipoint extension of this Ward identity based on our perturbative results.

\subsection{Five-point functions}
\label{subsec:5pt}

We now move to the more interesting case of five-point functions. We start by reviewing the kinematics specific to this case, before giving some examples of correlators for low-lying operators.

\subsubsection{Kinematics}
\label{subsubsec:5kinematics}

In this section, we review explicitly the kinematics introduced at the end of section \ref{subsec:correlators}. Here there are \textit{two} spacetime cross-ratios, which are defined as follows:
\begin{equation}
\chi_1 = \frac{\tau_{12} \tau_{45}}{\tau_{14} \tau_{25}}\,, \qquad \chi_2 = \frac{\tau_{13} \tau_{45}}{\tau_{14} \tau_{35}}\,.
\end{equation}
On the other hand, there are \textit{five} $R$-symmetry cross-ratios, for which we choose the basis to be\footnote{Note that we drop the subscript of $t_{ij}$,  since in this case there is only one $R$-symmetry cross-ratio of this kind.}
\begin{gather}
r_1 = \frac{(u_1 \cdot u_2)(u_4 \cdot u_5)}{(u_1 \cdot u_4)(u_2 \cdot u_5)}\,, \qquad s_1 = \frac{(u_1 \cdot u_5)(u_2 \cdot u_4)}{(u_1 \cdot u_4)(u_2 \cdot u_5)}\,, \notag \\
r_2 = \frac{(u_1 \cdot u_3)(u_4 \cdot u_5)}{(u_1 \cdot u_4)(u_3 \cdot u_5)}\,, \qquad s_2 = \frac{(u_1 \cdot u_5)(u_3 \cdot u_4)}{(u_1 \cdot u_4)(u_3 \cdot u_5)}\,, \notag \\
t = \frac{(u_1 \cdot u_5)(u_2 \cdot u_3)(u_4 \cdot u_5)}{(u_1 \cdot u_4)(u_2 \cdot u_5)(u_3 \cdot u_5)}\,.
\end{gather}
Using these cross-ratios the correlators can be expressed in terms of \textit{$R$-symmetry channels}. The number of channels depends on the scaling dimensions of the external operators (see table \ref{table:Rsymmchannels} for some examples).  Understanding this number is only a combinatorial matter. Let's take the first example of five-point that we are going to analyze in the following: $\vev{\Op_1 \Op_1 \Op_1 \Op_1 \Op_2}$. We have to consider all the possible combinations of the $R$-symmetry vectors $u$: $u_1$, $u_2$, $u_3$, $u_4$, $u_5$ and again $u_5$, since the last operator is $\Op_2$. We have to remember that we cannot ``pair'' the $R$-symmetry vectors associated to the same operator,  in this case the $u_5$ vectors. This is ensured by the properties of the $u$ vectors introduced below (\ref{eq:defectoperators}). Therefore we can make in total 6 different combinations that we write below:
\begin{equation}
\begin{aligned}
&(u_1\cdot u_2)(u_3\cdot u_5)(u_4\cdot u_5)\,, \\
&(u_1\cdot u_3)(u_2\cdot u_5)(u_4\cdot u_5)\,, \\
&(u_1\cdot u_4)(u_2\cdot u_5)(u_3\cdot u_5)\,,  \\
&(u_1\cdot u_5)(u_2\cdot u_3)(u_4\cdot u_5)\,,  \\
&(u_1\cdot u_5)(u_2\cdot u_4)(u_3\cdot u_5)\,, \\
&(u_1\cdot u_5)(u_2\cdot u_5)(u_3\cdot u_4)\,.
\end{aligned}
\end{equation}
This strategy can be easily implemented to compute the $R$-symmetry channels of all the correlators. However it is not straightforward to obtain a formula for the number of channels in the most generic case.  We thus determined it for the special case in which all external dimensions are equal to one,  where this number is reproduced by the double factorial: $(n-1)!! = 1 \cdot 3 \cdot 5 \cdot \ldots \cdot (n-1)$. Here $n$ is the (even) number of operators in the correlation function, and for example, the six-point function $\vev{\Op_1 \Op_1 \Op_1 \Op_1 \Op_1 \Op_1}$ consists of $5!!=1 \cdot 3 \cdot 5=15$ $R$-symmetry channels.

For simplicity, in this section we will focus on the correlation functions having up to $10$ channels only. In general these correlators take the form
\begin{align}
\Am_{\Delta_1 \Delta_2 \Delta_3 \Delta_4 \Delta_5} =& F_0 + \frac{\chi_1^2}{r_1} F_1 + \frac{r_2}{r_1} \frac{\chi_1^2}{\chi_2^2} F_2 + \frac{s_1}{r_1} \frac{\chi_1^2}{(1-\chi_1)^2} F_3 +  \frac{s_2}{r_1} \frac{\chi_1^2}{(1-\chi_2)^2} F_4 \notag \\
& + \frac{t}{r_1} \frac{\chi_1^2}{(\chi_1-\chi_2)^2} F_5 +  \frac{s_1}{r_1 s_2} \frac{\chi_1^2 (1-\chi_2)^2}{(1-\chi_1)^2} F_6 +  \frac{t}{r_1 s_2} \frac{\chi_1^2 (1-\chi_2)^2}{(\chi_1-\chi_2)^2} F_7 \notag \\
& +  \frac{r_2 s_1}{r_1 s_2} \frac{\chi_1^2 (1-\chi_2)^2}{\chi_2^2 (1-\chi_1)^2} F_8 +  \frac{r_2 t}{r_1 s_2} \frac{\chi_1^2 (1-\chi_2)^2}{\chi_2^2 (\chi_1-\chi_2)^2} F_9\,,
\end{align}
where we suppressed the dependency on the spacetime cross-ratios for compactness, i.e. $F_j := F_j (\chi_1, \chi_2)$. The $R$-symmetry channels have the following perturbative expansion
\begin{equation}
F_{j} = \lambda^{\frac{l}{2}}
\sum_{n=0}^{\infty} \lambda^{k}
F^{(k)}_{j}\, ,\qquad  l:= \sum_{i=1}^5\Delta_{i}\, .
\end{equation}
The prefactor for each channel is chosen such that it becomes $1$ in the topological limit. At finite $N$,  all the channels would be present at any loop order but in the planar limit $N \to \infty$ many channels do not contribute at least for the NLO computations, as we will soon see.

With these definitions the topological sector corresponds to:
\begin{equation}
\Am_{\Delta_1 \Delta_2 \Delta_3 \Delta_4 \Delta_5}
\raisebox{-2ex}{$\Biggr|$}_{\raisebox{2ex}{${\begin{subarray}{l} r_i \to  \chi_i^2 \\
 s_i \to  (1-\chi_i)^2
 \\ t \to (\chi_1 - \chi_2)^2\end{subarray}}$}}
 = \text{constant}\,.
\end{equation}
The constant on the right-hand side can be determined either by using supersymmetric localization techniques \cite{Giombi:2018qox,Giombi:2018hsx} or by pinching the full correlator up to two- or three-point functions,  in order to compare it to the NLO results given in \eqref{eq:twopt} and \eqref{eq:threept}. 

\begin{table}[H]
\centering
\caption{Number of $R$-symmetry channels for different five-point functions, obtained by pinching $n$-point functions $\vev{\Op_1 \ldots \Op_1}$ from $n=6$ to $n=12$. }
\begin{subtable}[c]{0.5\textwidth}
\centering
\begin{tabular}{cc}
\hline
 $\Delta_1$, $\Delta_2$, $\Delta_3$, $\Delta_4$, $\Delta_5$ & channels \\
\hline
\vspace{0.267cm}
$1,1,1,1,2$ & 6  \\
$1,1,1,1,4$ & 1  \\
$1,1,1,2,3$ & 6  \\
\vspace{0.26cm}
$1,1,2,2,2$ & 10\\
$1,1,1,2,5$ & 1  \\
$1,1,1,3,4$ & 6\\
$1,1,2,2,4$ & 6 \\
$1,1,2,3,3$ & 10  \\
$1,2,2,2,3$ & 15 \\
$2,2,2,2,2$ & 22 \\ \hline
\end{tabular}
\end{subtable}
\hspace{-2.2cm}
\begin{subtable}[c]{0.5\textwidth}
\centering
\begin{tabular}{cc}
\hline
$\Delta_1$, $\Delta_2$, $\Delta_3$, $\Delta_4$, $\Delta_5$ & channels \\
\hline
$1,1,1,3,6$ & 1 \\
$1,1,2,2,6$ & 1 \\
$1,1,1,4,5$ & 6 \\
$1,1,2,3,5$ & 6\\
$1,2,2,2,5$ & 6 \\
$1,1,2,4,4$ & 10 \\
$1,1,3,3,4$ & 10 \\
$1,2,2,3,4$ & 15 \\
$2,2,2,2,4$ & 21 \\
$1,2,3,3,3$ & 21\\
$2,2,2,3,3$ & 29 \\ \hline
\end{tabular}
\end{subtable}
\label{table:Rsymmchannels}
\end{table}

\subsubsection{$\vev{\Op_1 \Op_1 \Op_1 \Op_1 \Op_2}$}
\label{subsubsec:11112}

We will start by considering the simplest pinching case, which is obtained by bringing the last two operators of the six-point function $\vev{\Op_1 \Op_1 \Op_1 \Op_1 \Op_1 \Op_1}$ together, i.e. by taking the limits
\begin{equation}
(u_6,\tau_6) \to (u_5, \tau_5)\,.
\end{equation}
We choose the superconformal prefactor to be:
\begin{equation}
\Km_{11112} = \frac{(u_1 \cdot u_2)(u_3 \cdot u_5)(u_4 \cdot u_5)}{\tau_{12}^2 \tau_{35}^2 \tau_{45}^2}\,.
\end{equation}

This correlator consists in principle of \textit{six} $R$-symmetry channels, but at leading order we find that only three channels do not vanish:
\begin{align*}
F_0^{(0)} = F_2^{(0)} = F_5^{(0)} = \frac{1}{512 \pi^6}\,, \qquad F_j^{(0)} = 0 \text{ otherwise.}
\end{align*}

At next-to-leading order, almost all the channels are present and we obtain the following contributions:
\begin{align*}
F_0^{(1)} =& - \frac{1}{12288 \pi^6} + \frac{1}{4096 \pi^8 (\chi_2 - \chi_1)} \left( \ell(\chi_1 , \chi_2)+2(\chi_2 - \chi_1) \left( L_R \left( \frac{\chi_1 - \chi_2}{\chi_1} \right)+\frac{i \pi}{2} \log \frac{\chi_1}{\chi_2} \right) \right)\,,  \\
F_1^{(1)} =& 0\,, \\
F_2^{(1)} =& - \frac{\chi_2}{4096 \pi^8 \chi_1 (\chi_2 - \chi_1)} \ell(\chi_1 , \chi_2)\,, \\
F_3^{(1)} =& \frac{\chi_2}{4096 \pi^8 \chi_1 (\chi_2 - \chi_1)} (\ell(1-\chi_1,1-\chi_2) + i \pi (\chi_2 - \chi_1))\,, \\
F_4^{(1)} =& - \frac{1}{12288 \pi^6} - \frac{1}{4096 \pi^8 (\chi_2 - \chi_1)} \left( \ell(\chi_1 , \chi_2) \phantom{\frac{1}{2}} \right. \\
& \qquad\qquad\qquad\qquad\qquad\qquad \left. - (\chi_2 - \chi_1) \left( L_R \left( \frac{\chi_1 - \chi_2}{1-\chi_2} \right) - i\pi \left( 1 + \log \frac{1-\chi_1}{1-\chi_2} \right) \right) \right)\,, \\
F_5^{(1)} =& - \frac{5}{24576 \pi^6} - \frac{1}{4096 \pi^8 \chi_1 (1 - \chi_2)} \left( \chi_1 \ell(1-\chi_1 , 1-\chi_2) - (1-\chi_2) \ell(\chi_1 , \chi_2) \phantom{\frac{1}{2}} \right. \\
& \qquad\qquad\qquad \left. + \chi_1 (1-\chi_2) \left( \text{Li}_2 \left( \frac{1-\chi_1}{\chi_2 - \chi_1} \right) - \text{Li}_2 \left( - \frac{1-\chi_2}{\chi_2 - \chi_1} \right) - 2 L_R \left( \frac{\chi_1}{\chi_1 - \chi_2} \right) \right) \right. \\
& \qquad\qquad\qquad\qquad\qquad\qquad \left. + i\pi \chi_1 \left( (\chi_2 - \chi_1) + (1-\chi_2) \log \left( - \frac{\chi_2 (1-\chi_1)}{(\chi_1 - \chi_2)^2} \right) \right) \right)\,.
\end{align*}
In the result above, we used the Rogers dilogarithm, defined in (\ref{eq:Rogers}) and the function $\ell(\chi_1, \chi_2)$, defined in (\ref{eq:smalll}).

Some checks can be performed on this result. First the channels are individually finite, as expected for correlators of protected operators. In fact, the pinching of operators produces divergences, but they cancel again when summing up the different contributions. It is possible to further pinch the operators of the five-point function in order to produce e.g. four- and three-point functions. In particular we checked that pinching $\vev{\Op_1 \Op_1 \Op_1 \Op_1 \Op_2}$ accordingly matches the known results for $\vev{\Op_1 \Op_1 \Op_1 \Op_3}$, $\vev{\Op_1 \Op_1 \Op_2 \Op_2}$ and $\vev{\Op_1 \Op_2 \Op_3}$. 

Curiously, the channel $F_4^{(1)}$ has an additional constant term compared to the other channels. We note also that the correlator given above seems to be in fact the top component of a \textit{family} of correlators, namely $\vev{\Op_1 \Op_1 \Op_1 \Op_k \Op_{k+1}}$, for which only the prefactor changes channelwise.  We collected additional correlators of this family in the ancillary \textsc{Mathematica} notebook. This classification into families of correlators is expected to hold only at next-to-leading order and in the planar limit.

\subsubsection{$\vev{\Op_1 \Op_1 \Op_2 \Op_2 \Op_4}$}
\label{subsubsec:11224}

We turn now our attention to a different family of operators, namely the ones which are similar or equal to the case $\vev{\Op_1 \Op_1 \Op_2 \Op_2 \Op_4}$. This is obtained by starting with the ten-point correlator of $\Op_1$'s and pinching accordingly.  We choose in this case the superconformal prefactor to be
\begin{equation}
\Km_{11224} = \frac{(u_1 \cdot u_2)(u_3 \cdot u_5)^2(u_4 \cdot u_5)^2}{\tau_{12}^2 \tau_{35}^4 \tau_{45}^4}\,.
\end{equation}

This five-point function also consists of six $R$-symmetry channels, and at leading order, as for $\vev{\Op_1 \Op_1 \Op_1 \Op_1 \Op_2}$, we find that only three channels contribute:
\begin{align*}
F_0^{(0)} = F_2^{(0)} = F_5^{(0)} = \frac{1}{32768 \pi^{10}}\,, \qquad F_j^{(0)} = 0 \text{ otherwise.}
\end{align*}

Surprisingly, at next-to-leading order there is no additional channel appearing and we have
\begin{align*}
F_0^{(1)} =& - \frac{1}{786432 \pi^{10}} + \frac{1}{262144 \pi^{12} (\chi_2 - \chi_1)} \left( \ell(\chi_1 , \chi_2) \phantom{\frac{1}{2}} \right. \\
& \qquad\qquad\qquad\qquad\qquad\qquad\qquad\qquad \left. +2(\chi_2 - \chi_1) \left( L_R \left( \frac{\chi_1 - \chi_2}{\chi_1} \right) + \frac{i \pi}{2} \log \frac{\chi_1}{\chi_2} \right) \right)\,,  \\
F_2^{(1)} =& - \frac{\chi_2}{262144 \pi^{12} \chi_1 (\chi_2 - \chi_1)} \ell(\chi_1 , \chi_2)\,, \\
F_5^{(1)} =& \frac{1}{262144 \pi^{12} \chi_1} \left( \ell(\chi_1, \chi_2) + 2\chi_1 \left( L_R \left( \frac{\chi_1}{\chi_1 - \chi_2} \right) + \frac{i\pi}{2} \log \frac{\chi_2 - \chi_1}{\chi_2} \right) \right) \,, \\
F_j ^{(1)}=& \,0 \text{ otherwise.}
\end{align*}
We note in particular that $F_0^{(1)}$ and $F_2^{(1)}$ are identical to the case $\vev{\Op_1 \Op_1 \Op_1 \Op_1 \Op_2}$ of the previous section, up to an overall prefactor. Again the divergences generated by pinching cancel each other as in the previous section.

Additional correlators of this type are included in the ancillary \textsc{Mathematica} notebook. These are $\vev{\Op_1 \Op_1 \Op_2 \Op_3 \Op_5}$, $\vev{\Op_1 \Op_1 \Op_2 \Op_4 \Op_6}$ and $\vev{\Op_1 \Op_1 \Op_3 \Op_3 \Op_6}$ (the two latter being exactly identical, numerical prefactor included). Interestingly, although the correlator $\vev{\Op_1 \Op_1 \Op_3 \Op_3 \Op_4}$ consists in principle of ten $R$-symmetry channels, we find that it belongs in fact to the same family as $\vev{\Op_1 \Op_1 \Op_2 \Op_2 \Op_4}$, as the additional channels do not contribute at this order.

\subsubsection{$\vev{\Op_1 \Op_1 \Op_2 \Op_2 \Op_2}$}
\label{subsubsec:11222}

We end our review of five-point functions by considering one case where \textit{ten} $R$-symmetry channels contribute. Here the conformal prefactor is chosen to be
\begin{equation}
\Km_{11222} = \frac{(u_1 \cdot u_2)(u_3 \cdot u_4)(u_3 \cdot u_5)(u_4 \cdot u_5)}{\tau_{12}^2 \tau_{34}^2 \tau_{35}^2 \tau_{45}^2}\,,
\end{equation}
and at leading order we obtain
\begin{align*}
F_0^{(0)} = F_4^{(0)} = F_5^{(0)} = F_9^{(0)} = \frac{1}{4096 \pi^8}\,, \qquad F_j^{(0)} = 0 \text{ otherwise}\,,
\end{align*}
similarly to the other cases.

The NLO computation results into
\begin{align*}
F_0^{(1)} =& \frac{1}{32768 \pi^{10} (\chi_2 - \chi_1)} \left( \ell(\chi_1 , \chi_2)+2(\chi_2 - \chi_1) \left( L_R \left( \frac{\chi_1 - \chi_2}{\chi_1} \right) + \frac{i \pi}{2} \log \frac{\chi_1}{\chi_2} \right) \right)\,,  \\
F_2^{(1)} =& - \frac{\chi_2}{32768 \pi^{10} \chi_1 (\chi_2 - \chi_1)} \ell(\chi_1 , \chi_2)\,, \\
F_3^{(1)} =& \frac{\chi_2}{32768 \pi^{10} \chi_1 (\chi_2 - \chi_1)} (\ell(1-\chi_1,1-\chi_2) + i \pi (\chi_2 - \chi_1))\,, \\
F_4^{(1)} =& - \frac{1}{98304 \pi^{10}} - \frac{1}{32768 \pi^{10} (\chi_2 - \chi_1)} \left( \ell(\chi_1 , \chi_2) \phantom{\frac{1}{2}} \right. \\
& \qquad\qquad\qquad\qquad\qquad\qquad \left. - (\chi_2 - \chi_1) \left( L_R \left( \frac{\chi_1 - \chi_2}{1-\chi_2} \right) - i\pi \left( 1 + \log \frac{1-\chi_1}{1-\chi_2} \right) \right) \right)\,, \\
F_5^{(1)} =& \frac{1}{196608 \pi^8} + \frac{1}{32768 \pi^{10}} \left( \frac{\ell(\chi_1, \chi_2)}{\chi_1} - \frac{\ell(1-\chi_1, 1-\chi_2)}{1-\chi_2} +i\pi \frac{\chi_1 - \chi_2}{1-\chi_2} \right. \\
& \qquad\qquad\qquad \left. +2 L_R \left(\frac{\chi_1}{\chi_1 - \chi_2} \right) +2 L_R \left(- \frac{\chi_2}{1-\chi_2} \right) + \log \chi_2 + \frac{1-\chi_2}{\chi_2} \log (1-\chi_2) \right. \\
& \qquad\qquad\qquad \left.+ \text{Li}_2 \left( \frac{1-\chi_1}{1-\chi_2} \right) - \text{Li}_2 \left( \frac{\chi_1-\chi_2}{1-\chi_2} \right) + i\pi \log \frac{(1-\chi_1)(1-\chi_2)(\chi_2-\chi_1)}{\chi_2(1-\chi_2)} \right)\,, \\
F_7^{(1)} =& - \frac{1}{32768 \pi^{10}} \left( \frac{\log \chi_2}{1-\chi_2} + \frac{\log(1-\chi_2)}{\chi_2} \right)\,, \\
F_9^{(1)} =& -\frac{1}{196608 \pi^8} + \frac{1}{32768 \pi^{10}} \left( \frac{\chi_2}{1-\chi_2} \log \chi_2 + \log (1-\chi_2) \right. \\
& \qquad\qquad\qquad\qquad\qquad\qquad\qquad\qquad \left. - \text{Li}_2 \left( \frac{1}{\chi_2} \right) + \text{Li}_2 \left( - \frac{1-\chi_2}{\chi_2} \right) + i\pi \log \chi_2 \right)\,, \\
F_j ^{(1)}=& \,0 \text{ otherwise.}
\end{align*}
Note that the additional channels appearing, namely $F_7^{(1)}$ and $F_9^{(1)}$, depend only on the cross-ratio $\chi_2$. As it was the case in the previous section, the divergences appearing by pinching the operators cancel perfectly and this results into a finite correlator.

This family of correlators\footnote{Note however that the constant term in the channel $F_4$ of $\vev{\Op_1 \Op_1 \Op_2 \Op_2 \Op_2}$ is absent in the other correlators we looked at.} includes at least $\vev{\Op_1 \Op_1 \Op_2 \Op_3 \Op_3}$ and $\vev{\Op_1 \Op_1 \Op_2 \Op_4 \Op_4}$, which can be found in the ancillary \textsc{Mathematica} notebook.

\subsection{Six-point functions}
\label{subsec:6pt}

The method described in the previous subsection can be extended straightforwardly to six-point functions. But even if our recursion relation is very efficient at producing correlators,  it becomes increasingly difficult to express them in terms of conformal cross-ratios. In particular, there are now \textit{three} spacetime cross-ratios 
\begin{align}
\chi_1= \frac{\tau_{12}\tau_{56}}{\tau_{15}\tau_{26}}\,,\quad \chi_2= \frac{\tau_{13}\tau_{56}}{\tau_{15}\tau_{36}}\,,\quad \chi_3= \frac{\tau_{14}\tau_{56}}{\tau_{15}\tau_{46}}\,,
\end{align}
and \textit{nine} $R$-symmetry cross-ratios 
\begin{gather} \label{6pt_Rsymmch}
r_1 = \frac{(u_1 \cdot u_2)(u_5 \cdot u_6)}{(u_1 \cdot u_5)(u_2 \cdot u_6)}\,, \quad r_2 = \frac{(u_1 \cdot u_3)(u_5 \cdot u_6)}{(u_1 \cdot u_5)(u_3 \cdot u_6)}\,, \quad r_3 = \frac{(u_1 \cdot u_4)(u_5 \cdot u_6)}{(u_1 \cdot u_5)(u_4 \cdot u_6)}\,, \notag \\
s_1 = \frac{(u_1 \cdot u_6)(u_2 \cdot u_5)}{(u_1 \cdot u_5)(u_2 \cdot u_6)}\,, \quad s_2 = \frac{(u_1 \cdot u_6)(u_3 \cdot u_5)}{(u_1 \cdot u_5)(u_3 \cdot u_6)}\,, \quad s_3 = \frac{(u_1 \cdot u_6)(u_4 \cdot u_5)}{(u_1 \cdot u_5)(u_4 \cdot u_6)}\,, \notag\\
t_{12}= \frac{(u_1 \cdot u_6)(u_2 \cdot u_3)(u_5 \cdot u_6)}{(u_1 \cdot u_5)(u_2 \cdot u_6)(u_3 \cdot u_6)}\,,\quad t_{13}= \frac{(u_1 \cdot u_6)(u_2 \cdot u_4)(u_5 \cdot u_6)}{(u_1 \cdot u_5)(u_2 \cdot u_6)(u_4 \cdot u_6)}\,, \notag \\
t_{23}= \frac{(u_1 \cdot u_6)(u_3 \cdot u_4)(u_5 \cdot u_6)}{(u_1 \cdot u_5)(u_3 \cdot u_6)(u_4 \cdot u_6)}  \,.
\end{gather}

Looking at the number of $R$-symmetry channels,  we notice that we do not obtain six-point functions with six channels,  as in the case of five-point functions. Indeed, all six-point correlators have at least ten $R$-symmetry channels\footnote{Here we are ignoring extremal correlators, which always have only one $R$-symmetry channel.}. For this reason, and since they are lengthy,  in the following we are going to present explicitly only one particularly simple six-point function. Nonetheless, we collected in the \textsc{Mathematica} notebook a few other examples, namely $\vev{\Op_1 \Op_1 \Op_1 \Op_1 \Op_1 \Op_1}$, $\vev{\Op_1 \Op_1 \Op_1 \Op_1 \Op_1 \Op_3}$, $\vev{\Op_1 \Op_1 \Op_1 \Op_1 \Op_2 \Op_4}$, $\vev{\Op_1 \Op_1 \Op_1 \Op_1 \Op_3 \Op_5}$.

Similarly to the five-point case, we notice that some correlators share the same structure. For example $\vev{\Op_1 \Op_1 \Op_1 \Op_1 \Op_1 \Op_3}$, $\vev{\Op_1 \Op_1 \Op_1 \Op_1 \Op_2 \Op_4}$, $\vev{\Op_1 \Op_1 \Op_1 \Op_1 \Op_3 \Op_5}$ belong to the family $\vev{\Op_1 \Op_1 \Op_1 \Op_1 \Op_k \Op_{k+2}}$.

\subsubsection{$\vev{\Op_1 \Op_1 \Op_1 \Op_2 \Op_2 \Op_5}$}
\label{subsubsec:111225}

The example we consider is the correlator $\vev{\Op_1 \Op_1 \Op_1 \Op_2 \Op_2 \Op_5}$, that we obtain by pinching the last three operators in the twelve-point function $\vev{\Op_1 \Op_1 \dots \Op_1  \Op_1 }$ in the following way:
\begin{gather}
(u_{12},\tau_{12};u_{11},\tau_{11};u_{10},\tau_{10};u_9,\tau_9;u_8,\tau_8) \rightarrow (u_6,\tau_6)\,,\notag \\
(u_7,\tau_7;u_6,\tau_6) \rightarrow (u_5,\tau_5)\,,\quad (u_5,\tau_5) \rightarrow (u_4,\tau_4)\,.
\end{gather}
We choose the superconformal prefactor to be:
\begin{equation}
\Km_{111225} = \frac{(u_1 \cdot u_2)(u_3 \cdot u_6)(u_4 \cdot u_6)^2(u_5 \cdot u_6)^2}{\tau_{12}^2 \tau_{36}^2 \tau_{46}^4 \tau_{56}^4}\,.
\end{equation}

In analogy with the five-point functions, we use the $R$-symmetry cross-ratios (\ref{6pt_Rsymmch}) to express the correlator in terms of $R$-symmetry channels:
\begin{align}
\Am_{\Delta_1 \Delta_2 \Delta_3 \Delta_4 \Delta_5\Delta_6} =& F_0 + \frac{\chi_1^2}{r_1} F_1 + \frac{r_2}{r_1} \frac{\chi_1^2}{\chi_2^2} F_2 + \frac{r_3}{r_1} \frac{\chi_1^2}{\chi_3^2} F_3 +  \frac{s_1}{r_1} \frac{\chi_1^2}{(1-\chi_1)^2} F_4 \notag \\
& + \frac{s_2}{r_1} \frac{\chi_1^2}{(1-\chi_2)^2} F_5+  \frac{s_3}{r_1} \frac{\chi_1^2}{(1-\chi_3)^2} F_6+\frac{t_{12}}{r_1} \frac{\chi_1^2}{(\chi_1-\chi_2)^2} F_7 \notag \\
& +  \frac{t_{13}}{r_1} \frac{\chi_1^2}{(\chi_1-\chi_3)^2} F_8 +  \frac{t_{23}}{r_1} \frac{\chi_1^2}{(\chi_2-\chi_3)^2} F_9\,,
\end{align}
where again the dependency on the spacetime cross-ratios is suppressed for compactness. As before, the $R$-symmetry channels can be expanded perturbatively as
\begin{equation}
F_{j} = \lambda^{\frac{l}{2}}
\sum_{n=0}^{\infty} \lambda^{k}
F^{(k)}_{j}\, ,\qquad  l:= \sum_{i=1}^6\Delta_{i}\, .
\end{equation}

This correlator consists in principle of \textit{ten} $R$-symmetry channels, but at leading order we find that only four channels do not vanish:
\begin{align}
F_0^{(0)} = F_6^{(0)} = F_7^{(0)} =F_9^{(0)}= \frac{1}{262144  \pi^{12}}\,, \qquad F_j^{(0)} = 0 \text{ otherwise.}
\end{align}

At next-to-leading order, six channels are non-vanishing and we obtain the following contributions:
\begin{align}
F_0^{(1)} =& - \frac{1}{6291456 \pi^{12}} + \frac{1}{2097152\pi^{14} (\chi_2 - \chi_1)} \left( \ell(\chi_1 , \chi_2) \phantom{\frac{1}{2}} \right. \notag\\
& \qquad\qquad\qquad\qquad\qquad\qquad\qquad\qquad \left. +2(\chi_2 - \chi_1) \left( L_R \left( \frac{\chi_1 - \chi_2}{\chi_1} \right) + \frac{i \pi}{2} \log \frac{\chi_1}{\chi_2} \right) \right)\,,  \notag\\
F_2^{(1)}=&\frac{\chi_2}{2097152 \pi^{14} \,\chi_1\, (\chi_1-\chi_2)} \ell(\chi_1,\chi_2) \,,\notag\\
F_7^{(1)}=&\frac{1}{12582912\pi^{12}}-\frac{1}{2097152\pi^{14}}\left(-\frac{\ell(\chi_1,\chi_2)}{\chi_1}-\frac{\ell(\chi_1,\chi_2)+\ell(\chi_2,\chi_3)-\ell(\chi_1,\chi_3)}{\chi_3-\chi_2}\right. \notag\,\\
&\qquad\qquad-2L_R\left.\left(\frac{\chi_1}{\chi_1-\chi_2}\right)+\text{Li}_2\left(\frac{\chi_3-\chi_1}{\chi_2-\chi_1}\right)-\text{Li}_2\left(\frac{\chi_2-\chi_3}{\chi_2-\chi_1}\right)- i \pi\, \text{log}\, \frac{(\chi_2-\chi_1)^2}{\chi_2\,(\chi_3-\chi_1)} \right)\notag\,,\\
F_8^{(1)}=&-\frac{(\chi_1-\chi_3)}{2097152 \pi^{14} (\chi_1-\chi_2)(\chi_2-\chi_3)} \left(-\ell(\chi_1,\chi_2)+ \ell(\chi_1,\chi_3)-\ell(\chi_2,\chi_3) \right)\,,\notag\\
F_9^{(1)}=&\frac{1}{2097152 \pi^{14}}  \left(\frac{\ell(\chi_1,\chi_2)-\ell(\chi_1,\chi_3)+\ell(\chi_2,\chi_3)}{\chi_2-\chi_1} +2L_R\left(\frac{\chi_1-\chi_2}{\chi_3-\chi_2}\right)+ i \pi\, \text{log}\, \frac{\chi_3-\chi_2}{\chi_3-\chi_1} \right)\,, \notag\\
F_j ^{(1)}=&\, 0 \text{ otherwise.}
\end{align}
It is interesting to note that the channels $F_0^{(1)}$ and $F_2^{(1)}$ share the exact same structure as the $F_0^{(1)}$ and $F_2^{(1)}$ of the five-point correlator $\vev{\Op_1 \Op_1 \Op_1 \Op_1 \Op_2}$ of section \ref{subsubsec:11112}.

\subsection{A conjecture for multipoint Ward identities}
\label{subsec:conjectureWI}

We have computed other correlators up to $n=8$, which can be found in the ancillary notebook. From these results we found experimentally that \textit{all} our correlators are
annihilated by the following family of differential operators: 
\begin{equation}
\sum_{k=1}^{n-3} \left( \frac{1}{2} \partial_{\chi_k} + \alpha_k \partial_{r_k} - (1-\alpha_k) \partial_{s_k} \right) \, \Am_{\Delta_1 \ldots \Delta_n} \raisebox{-2ex}{$\Biggr 
|$}_{\raisebox{1.5ex}{$\begin{subarray}{l} r_i \to \alpha_i \chi_i \\
 s_i \to (1-\alpha_i)(1-\chi_i)
 \\ t_{ij} \to (\alpha_i-\alpha_j)(\chi_i-\chi_j)\end{subarray}$}} = 0\,,
\label{eq:WIn}
\end{equation}
with $\alpha_k$ being \emph{arbitrary} real numbers. Notice that these operators are a natural generalization of the differential operator that captures the Ward identity \eqref{eq:WI4} for four-point functions of half-BPS operators.  We then conjecture that these equations are a multipoint extension of the superconformal Ward identity satisfied by the four-point functions. 

Even though we obtained these equations from NLO correlators, we expect these identities to be also satisfied in the strong-coupling expansion. This regime is captured by a well-understood AdS dual \cite{Giombi:2017cqn}, which has been used to calculate planar correlators in the $\lambda \to \infty$ limit. These are given by simple Wick contractions of the fluctuations of the dual fundamental string, i.e. the leading, disconnected order corresponds to the generalized free-field expression, e.g. (4.5) in \cite{Giombi:2017cqn}. It is then easy to check that all our $n$-point functions also satisfy \eqref{eq:WIn} in the extreme strong-coupling limit. We have therefore three non-trivial data points: the first two orders at weak coupling, and the leading term at strong coupling. It is then reasonable to assume that the constraint \eqref{eq:WIn} is non-perturbative and valid at \emph{all} loop orders. Notice also that superconformal constraints are insensitive to gauge-theory quantum numbers, which means our identities should also hold for non-planar corrections. Indeed, we checked that this is the case for the first correction in the $N$ expansion, up to $n=8$.

However we should point out that our conjecture cannot represent the full set of superconformal constraints on the correlators. The reason is because our analysis of protected operators only focuses on the highest-weight component, and we are ignoring possible fermionic descendants. Working in a suitable superspace, it is known that for four-points the full superconformal correlator can be reconstructed from the highest weight, and so it is safe to set the fermions to zero. Starting with five-point and up, one expects nilpotent invariants\footnote{We thank Paul Heslop for discussions.}. In general, for generic $n$-point functions the Ward identities should be a collection of partial PDEs relating the components associated to each fermionic structure. The fact that we obtained a differential operator that only acts on the highest weight and still annihilates the correlator is unexpected. It would be nice to do a proper superspace analysis (similar to what was done in \cite{Chicherin:2015bza}) and prove that our experimental observation is indeed one of the constraints imposed by superconformal invariance. 

\endgroup
\section{Conclusions}
\label{sec:conclusions}

In this work we developed an efficient algorithm for computing multipoint correlation functions of protected operators in the $1d$ Maldacena-Wilson line CFT. These are new perturbative results for an interesting model that has received increased attention in recent years. Even though our correlators are interesting in their own right, they gave us an unexpected, more general, piece of information: multipoint Ward identities. After gathering a significant collection of correlators, we found experimentally that all of them are annihilated by a family of differential operators. Four-point superconformal Ward identities of the type studied here are known to be satisfied in several superconformal setups \cite{Liendo:2016ymz,Liendo:2015cgi,Dolan:2004mu}. Our differential operators are a natural generalization of these Ward identities to multipoint correlators. Even though we obtained them from a next-to-leading order analysis, we expect the result to be more general and valid at any loop order.

There are many interesting directions in which to further develop the techniques presented in this article. One unexplored aspect of the recursion relation given in \eqref{eq:recursionloop} is that it can  be used to calculate correlators of  \textit{unprotected} scalar operators. For example, at length $\Delta = 2$ we can build \cite{Correa:2018fgz}:
\begin{equation}
\Op^{ij}_A := \phi^i \phi^j - \phi^j \phi^i \,, \qquad i,j = 1, \ldots, 5\,, 
\end{equation}
which can be obtained by pinching two $\Delta = 1$ operators. In particular, the two-point function $\vev{\Op^{ij}_A (\tau_1) \Op^{kl}_A (\tau_2)}$ is easy to produce using the results of this paper and its scaling dimensions is
\begin{equation}
\Delta_A = 2 + \frac{\lambda}{4\pi^2} + \Op(\lambda^2)\,,
\end{equation}
which perfectly matches the results of \cite{Correa:2018fgz} for the supersymmetric case $\zeta = 1$.
It is straightforward to extend this analysis to higher leading-order dimensions, which means that through pinching of the fundamental field one can write recursion relations for anomalous dimensions of unprotected operators. The only missing ingredient in this discussion is the remaining scalar field $\phi^6$. The next natural target is to include  $\phi^6$ in our recursion relations, such that we have access to all single-trace scalar operators \cite{Barrat:2022wip}.
   
It would be very interesting to prove our conjectured multipoint Ward identities.
The usual method relies on superspace techniques and requires a careful analysis of possible nilpotent invariants. Techniques relevant for such an analysis have been developed in \cite{Chicherin:2015bza} for $\Nm=4$ SYM without defects. It would be interesting to adapt that work to our one-dimensional setup.  
We expect the supersymmetric completion of the Maldacena-Wilson loop operator of
\cite{Muller:2013rta,Beisert:2015jxa,Beisert:2015uda} to be of relevance here.
Further tests can be performed on the conjectured Ward identities in the string dual. For example, a formula was recently derived \cite{Bliard:2022xsm} for computing arbitrary contact diagrams involving external scalar operators in AdS$_2$, this can be used to check our identities for the leading order connected diagram at strong coupling. 

In addition,  a closed-form expression for leading-order contact interactions of identical scalars at strong coupling was recently found in Mellin space through an inherently one-dimensional Mellin formalism \cite{Bianchi:2021piu}.  Mellin space is known to offer an alternative, sometimes simpler, formulation for correlation functions.  It would be interesting to extend \cite{Bianchi:2021piu} to multipoint correlators, and connect that formalism with the results presented here. 

As discussed in the introduction, multipoint correlators are one of the long-term goals of the bootstrap program for CFTs. Recent developments include a theory for conformal blocks \cite{Buric:2020dyz,Buric:2021ywo}, and a light-cone analysis \cite{Antunes:2021kmm,Anous:2021caj}. Due to its simplicity however, $1d$ is usually a convenient starting point. Our results are explicit examples of dynamical multipoint correlators in a valid $1d$ defect CFT, which can be used as a testing ground for multipoint bootstrap techniques. They are also interesting in their own right, as they contain an infinite amount of CFT data which also includes non-protected operators.

Multipoint superconformal Ward identities is an underexplored subject. In principle, one could try to repeat the strategy presented here in more general setups. For bulk $\Nm=4$ SYM  a recursion relation has already been worked out for $n$-point functions of half-BPS operators at next-to-leading order \cite{Drukker:2008pi}. In the $4d$ case there will be more spacetime cross-ratios, and it would be interesting to see whether or not there exist superconformal Ward identities that only act on the scalar highest weight, as it seems to be the case in $1d$.

\acknowledgments

We are particularly grateful to G.~Bliard, L.~Corcoran, B.~Eden, V.~Forini, A.~Gimenez-Grau, P.~Heslop,  E.~Malek,  J.~Marschner, and L.~Quintavalle for useful discussions.
PL is supported by the DFG through the Emmy Noether research group ``The Conformal Bootstrap Program'' project number 400570283. JB and GP are funded by the Deutsche Forschungsgemeinschaft (DFG, German Research Foundation) -- Projektnummer\\ 417533893/GRK2575 ``Rethinking Quantum Field Theory''.

\appendix

\section{Insertion rules}
\label{sec:insertion}

In this appendix, we list the insertion rules used for computing the Feynman diagrams of section \ref{sec:correlators}. Those are derived from the action of $\mathcal{N}=4$ SYM in $4d$ Euclidean space, which is given by \eqref{eq:action}. Note that we consider $SU(N)$ as the gauge group and that we work in the large $N$ limit. The generators obey the following commutation relation:
\begin{equation}
[ T^a\,, T^b ] = i f^{abc}\, T^c,
\end{equation}
in which $f^{abc}$ are the structure constants of the $\mathfrak{su}(N)$ Lie algebra. The generators are normalized as
\begin{equation}
\tr T^a \tensor{T}{^b} = \frac{\delta^{ab}}{2}.
\end{equation}
Note that $f^{ab0} = 0$ and $\tr \tensor{T}{^a} = 0$. The (contracted) product of structure constants gives $f^{abc} f^{abc} = N (N^2 - 1) \sim N^3$, where the second equality holds in the large $N$ limit.

The only two-point insertion that we need is the self-energy of the scalar propagator at one loop, which is given by the following expression \cite{Erickson:2000af, Plefka:2001bu, Drukker:2008pi}:
\begin{align}
\propagatorSSEnotext\ &=\ \SSEone\ +\ \SSEtwo\ +\ \SSEthree\ +\ \SSEfour \notag \\
&= -2 g^4 N \delta^{ab} \delta_{ij} Y_{112}.
\label{eq:selfenergy}
\end{align}
The integral $Y_{112}$ is given in (\ref{eq:Y112}) and presents a logarithmic divergence.

We also require only one three-point insertion, which is the vertex connecting two scalar fields and one gauge field. It is easy to obtain from the action (\ref{eq:action}) and it reads
\begin{equation}
\vertexSSG\ = - g^4 f^{abc} \delta^{ij} \left( \partial_1 - \partial_2 \right)_\mu Y_{123}.
\label{eq:vertexSSG}
\end{equation}
The $Y$-integral is defined in (\ref{subeq:Y123}) and its analytical expression in $1d$ can be found in (\ref{eq:Y123}).

Another relevant vertex is the four-scalars coupling. Similarly to the three-vertex, it is straightforward to read the corresponding Feynman rule from the action and perform the Wick contractions in order to get
\begin{align}
\vertexSSSS &= -g^6 \left\lbrace f^{abe}f^{cde} \left( \delta_{ik}\delta_{jl} - \delta_{il}\delta_{jk} \right) + f^{ace}f^{bde} \left( \delta_{ij}\delta_{kl} - \delta_{il}\delta_{jk} \right) \right. \notag \\[-1.5em]
& \left. \qquad \qquad \qquad + f^{ade}f^{bce} \left( \delta_{ij}\delta_{kl} - \delta_{ik}\delta_{jl} \right) \right\rbrace X_{1234}.
\label{eq:vertexSSSS}
\end{align}
The $X$-integral can be found in (\ref{eq:X1234}). 

There is one more sophisticated four-point insertion that we require, which reads
\begin{equation}
\Hinsertone\ = g^6 \left\lbrace \delta_{ik} \delta_{jl} f^{ace} f^{bde} I_{13} I_{24} F_{13,24} + \delta_{il} \delta_{jk} f^{ade} f^{bce} I_{14} I_{23} F_{14,23} \right\rbrace\,.
\label{eq:Hinsertone}
\end{equation}
with $I_{ij}$ the propagator function defined in (\ref{eq:I12}) and $F_{ij,kl}$ as defined in (\ref{subeq:F1324}). An analytical expression for $F_{ij,kl}$ in terms of $X$- and $Y$-integrals is given in (\ref{eq:FXYidentity}).
\section{Integrals}
\label{sec:integrals}

In this appendix, we give some detail about the integrals used in this work.

\subsection{Standard integrals}
\label{subsec:standardintegrals}

In the computation of the Feynman diagrams at next-to-leading order, we encounter three-, four- and five-point massless Feynman integrals, which we define as follows:
\begin{subequations}
\begin{gather}
Y_{123} := \int d^4 x_4\ I_{14} I_{24} I_{34}\,, \label{subeq:Y123} \\
X_{1234} := \int d^4 x_5\ I_{15} I_{25} I_{35} I_{45}\,, \label{subeq:X1234} \\
H_{13,24} := \int d^4 x_{56}\ I_{15} I_{35} I_{26} I_{46} I_{56}\,, \label{subeq:H1324}
\intertext{with $I_{ij}$ the propagator function defined in (\ref{eq:I12}). In the last expression we have defined $d^4 x_{56} := d^4 x_5\ d^4 x_6$ for brevity. The letter assigned to each integral is evocative of the drawing of the propagators. Another expression which is encountered is the following:}
F_{13,24} := \frac{\left( \partial_1 - \partial_3 \right) \cdot \left(\partial_2 - \partial_4 \right)}{I_{13}I_{24}} H_{13,24}\,. \label{subeq:F1324}
\end{gather}
\end{subequations}
The notation presented above is standard and has already been used in e.g. \cite{Beisert:2002bb, Drukker:2008pi}. The three- and four-point massless integrals in Euclidean space are conformal and have been solved analytically (see e.g. \cite{tHooft:1978jhc, Usyukina:1992wz} and \cite{Drukker:2008pi,Kiryu:2018phb} for the modern notation). In $1d$ the $X$-integral is given by
\begin{equation}
\frac{X_{1234}}{I_{13} I_{24}} = - \frac{1}{8 \pi^2} \frac{\ell(\chi, 1)}{\chi(1-\chi)}\,,  \qquad\qquad \chi^2 := \frac{\tau_{12}^2 \tau_{34}^2}{\tau_{13}^2 \tau_{24}^2}\,,
\label{eq:X1234}
\end{equation}
with $\ell(\chi_1,\chi_2)$ defined in (\ref{eq:smalll}).
The $Y$-integral can easily be obtained from this expression by taking the following limit:
\begin{align}
Y_{123} &= \lim_{x_4 \to \infty} (2\pi)^2 x_4^2\ X_{1234} \notag \\
&= \frac{I_{12}}{8\pi^2} \left( \frac{\tau_{12}}{\tau_{23}} \log |\tau_{13}| + \frac{\tau_{12}}{\tau_{31}} \log |\tau_{23}| + \frac{\tau_{12}^2}{\tau_{23}\tau_{31}} \log |\tau_{12}| \right)\,.
\label{eq:Y123}
\end{align}

The $H$-integral seems to have no known closed form so far, but (\ref{subeq:F1324}) can fortunately be reduced to a sum of $Y$- and $X$-integrals in the following way \cite{Beisert:2002bb}:
\begin{align}
F_{13,24} &= \frac{X_{1234}}{I_{12}I_{34}} - \frac{X_{1234}}{I_{14}I_{23}} + \left( \frac{1}{I_{14}} - \frac{1}{I_{12}} \right) Y_{124} + \left( \frac{1}{I_{23}} - \frac{1}{I_{34}} \right) Y_{234} \notag \\
& \qquad \qquad \qquad \qquad \qquad + \left( \frac{1}{I_{23}} - \frac{1}{I_{12}} \right) Y_{123} + \left( \frac{1}{I_{14}} - \frac{1}{I_{34}} \right) Y_{134}\,.
\label{eq:FXYidentity}
\end{align}

The integrals given above also appear in their respective pinching limits, i.e. when two external points are brought close to each other. The integrals simplify greatly in this limit, but they exhibit a logarithmic divergence which is tamed by using point-splitting regularization. For the $Y$-integral we define
\begin{equation*}
Y_{122} := \lim_{x_3 \to x_2} Y_{123}\,, \qquad \qquad \qquad \lim_{x_3 \to x_2} I_{23} := \frac{1}{(2\pi)^2 \epsilon^2}\,.
\end{equation*}
%
Inserting this in (\ref{eq:Y123}) and expanding up to order $\mathcal{O} (\log \epsilon^2)$, we obtain
\begin{equation}
\diagYonetwotwo\ := Y_{112} = Y_{122} = - \frac{I_{12}}{16 \pi^2} \left( \log \frac{\epsilon^2}{\tau_{12}^2} - 2 \right)\,.
\label{eq:Y112}
\end{equation}
This result coincides with the expression given in e.g. \cite{Drukker:2008pi}.

For completion, we also give the pinching limit of the $X$- and $F$-integrals. The first one reads
\begin{equation}
\diagXoneonetwothree\ := X_{1123} = - \frac{I_{12} I_{13}}{16 \pi^2} \left( \log \frac{\epsilon^2 \tau_{23}^2}{\tau_{12}^2 \tau_{13}^2} - 2 \right)\,,
\label{eq:X1123}
\end{equation}
which is again the same as in \cite{Drukker:2008pi}.

Finally, the pinching limit $\tau_2 \to \tau_1$ of the $F$-integral gives
\begin{align}
F_{13,14} &= F_{14,13} = - F_{13,41} \notag \\
& = -\frac{X_{1134}}{I_{13} I_{14}} + \frac{Y_{113}}{I_{13}} + \frac{Y_{114}}{I_{14}} + \left(\frac{1}{I_{13}} + \frac{1}{I_{14}} - \frac{2}{I_{34}} \right) Y_{134}\,.
\label{eq:F1314}
\end{align}

\subsection{$T$-integrals}
\label{subsec:Tintegrals}

In presence of the line, there is a new type of integral arising in addition to the bulk integrals of the previous appendix. We denote this integral by $T_{ij;kl}$\footnote{This class of integrals also appears in \cite{Kiryu:2018phb}, where they are defined slightly differently and labelled as $B_{ij;kl}$.}, defined as
\begin{equation}
T_{ij;kl} := \partial_{ij} \int_{\tau_k}^{\tau_l} d\tau_m\, \epsilon(ijm)\, Y_{ijm}\,,
\label{eq:Tdef}
\end{equation}
where $\epsilon(ijk)$ encodes the change of sign due the path ordering and is defined as
\begin{equation}
\epsilon(ijk) := \text{sgn}\, \tau_{ij}\, \text{sgn}\, \tau_{ik}\, \text{sgn}\, \tau_{jk}\,.
\label{eq:epsdef}
\end{equation}

When the range of integration is the entire line, the integral is easy to perform and results in
\begin{equation}
T_{ij;(-\infty)(+\infty)} = - \frac{I_{ij}}{12}\,.
\label{eq:T1}
\end{equation}
In the case where $(i,j) = (k,l)$ it gives
\begin{equation}
T_{ij;ij} = \frac{I_{ij}}{12} \,.
\label{eq:T2}
\end{equation}

Let us now review some relations satisfied by the $T$-integrals. The following identity can be used in order to "swap" the limits of integration:
\begin{equation}
\left. T_{jk;il} \right|_{i<j<k<l} = -\frac{I_{jk}}{12} - T_{jk;li}\,,
\label{eq:T3}
\end{equation}
where the integration range $(li)$ on the right-hand side has to be understood as the union of segments $(l,+\infty) \cup (-\infty,i)$.

There also exist another relevant combination for the computations at one loop relating the $T$- and $Y$-integrals:
\begin{equation}
I_{ik} T_{jk;ki} + I_{jk} T_{ik;jk} = - \frac{I_{ik} I_{jk}}{12} +I_{ik} I_{jk}\left(\frac{1}{I_{ik}} + \frac{1}{I_{jk}}-\frac{2}{I_{ij}} \right)Y_{ijk}\,.
\label{eq:T5}
\end{equation}

In general the integrals can be performed explicitly for the different possible orderings of the $\tau$'s, and here we give the results assuming $\tau_1 < \tau_2 < \tau_3 < \tau_4$:
\begin{subequations}
\begin{gather}
T_{12;34} = \frac{1}{32 \pi^4 \tau_{12}^2} \left( 4 L_R \left( \frac{\tau_{12}}{\tau_{14}} \right) - 4 L_R \left( \frac{\tau_{12}}{\tau_{13}} \right) - C_{123} + C_{124} \right)\,, \\
T_{34;12} = \frac{1}{32 \pi^4 \tau_{34}^2} \left( 4 L_R \left( \frac{\tau_{34}}{\tau_{14}} \right) - 4 L_R \left( \frac{\tau_{34}}{\tau_{24}} \right) - C_{341} + C_{342} \right)\,, \\
T_{14;23} = \frac{1}{32 \pi^4 \tau_{14}^2} \left( 4 L_R \left( \frac{\tau_{24}}{\tau_{14}} \right) - 4 L_R \left( \frac{\tau_{34}}{\tau_{14}} \right) - C_{412} - C_{143} \right)\,, \\
T_{23;41} = \frac{1}{32 \pi^4 \tau_{23}^2} \left( -4 L_R \left( \frac{\tau_{23}}{\tau_{13}} \right) - 4 L_R \left( \frac{\tau_{23}}{\tau_{24}} \right) - C_{234} - C_{123} \right)\,,
\end{gather}
\label{eq:Tint}
\end{subequations}
where we have defined the following help function:
\begin{equation}
C_{ijk} := - 32 \pi^4 \tau_{ij} (\tau_{ik} + \tau_{jk}) Y_{ijk}\,,
\end{equation}
and where the Rogers dilogarithm $L_R(x)$ is defined in \eqref{eq:Rogers}.

It is easy to take pinching limits of the integrals given above. For example, we can have
\begin{equation}
T_{12;23} = \frac{1}{32 \pi^4 \tau_{12}^2} \left( 4 L_R \left( \frac{\tau_{12}}{\tau_{13}} \right) - \frac{2 \pi^2}{3} + C_{123} \right) + Y_{112}\,,
\end{equation}
using the fact that $L_R(1) = \frac{\pi^2}{6}$. All the other pinching limits can be performed in the same way.

\bibliography{./auxi/Notes.bib}
\bibliographystyle{./auxi/JHEP}

\end{document}